\documentclass[aps,prd,preprintnumbers,showpacs,showkeys,nofootinbib,
superscriptaddress,fleqn,floatfix,tightenlines,10pt]{revtex4-1}
\usepackage{amsmath,amsfonts,amssymb,amscd,amsxtra,amsthm}
\usepackage{graphicx}  % Include figure files
\usepackage{epstopdf}
\usepackage{dcolumn}  % Align table columns on decimal point 
\usepackage{bm}          % bold math
\usepackage{slashed}
\usepackage{cancel}
\usepackage{float}
\usepackage{mathtools}
\usepackage{amsbsy}
\usepackage{amstext}

\usepackage[utf8]{inputenc} 
\usepackage{booktabs} 
\usepackage[normalem]{ulem} % \sout{old text} for strikeout
\usepackage[dvipsnames]{xcolor} % For blue in-text comments and
\usepackage{tabularx}
\usepackage{enumitem}  
\usepackage{array} 
\usepackage{slashed}
\usepackage{tikz}
\usepackage{float}
\usepackage{multirow}
\usepackage{cancel}
\renewcommand\sout{\bgroup \color{red} \ULdepth=-.5ex \ULset}

\makeatletter

\begin{document}  
\preprint{INHA-NTG-05/2021}
\title{Transverse charge distributions of the nucleon and their Abel
  images} 
%--------------------------------------------------
%--------------------------------------------------
\author{June-Young Kim}
\email[E-mail: ]{Jun-Young.Kim@ruhr-uni-bochum.de}
\affiliation{Institut f\"ur Theoretische Physik II, Ruhr-Universit\"at
  Bochum, D-44780 Bochum, Germany}

\author{Hyun-Chul Kim}
\email[E-mail: ]{hchkim@inha.ac.kr}
\affiliation{Department of Physics, Inha University, Incheon 22212,
Republic of Korea}
\affiliation{School of Physics, Korea Institute for Advanced Study
  (KIAS), Seoul 02455, Republic of Korea}

%--------------------------------------------------
\date{\today}
\begin{abstract}
We investigate the two-dimensional transverse charge distributions of
the transversely polarized nucleon. As the longitudinal momentum
($P_z$) of the nucleon increases, the electric dipole moment is
induced, which causes the displacement of the transverse charge and
magnetization distributions of the nucleon. The induced dipole moment
of the proton reaches its maximum value at around $P_z \approx 3.2$
GeV due to the kinematical reason. We also investigate how the Abel
transformations map the three-dimensional charge and magnetization 
distributions in the Breit frame onto the transverse charge and
magnetization ones in the infinite momentum frame. 
\end{abstract}
\maketitle
%--------------------------------------------------

\section{Introduction}
The electromagnetic (EM) structure of the nucleon has been one of the
most important issues well over decades since Hofstadter's
experiments~\cite{Hofstadter:1956qs} (see a recent
review~\cite{Punjabi:2015bba} and references therein). The EM form
factors of the nucleon, which can be measured by electron-nucleon
scattering, provide information on the three-dimensional (3D) charge
and magnetization distributions for the nucleon in the Breit frame
(BF). They have traditionally been obtained by the 3D 
Fourier transform~\cite{Ernst:1960zza, Sachs:1962zzc}. However, this 
interpretation is true only when the size of a system is much larger
than the Compton wavelength~($\lambdabar = \hbar/Mc$). This means that
the definitions of the 3D charge and magnetization distributions are 
completely valid for nonrelativistic particles such as atoms and
nuclei. On the other hand, the charge radius of the nucleon is
approximately four times larger than the Compton wavelength, which
indicates that the relativistic corrections are about 20~\% for the
distributions and about 10~\% for the charge and magnetic radii of the
nucleon. This indicates that the nucleon is \emph{per se} a
relativistic particle. In fact, Yennie \emph{et al.} raised this point
already many years ago~\cite{Yennie:1957}. Recently, the criticisms on
the 3D distributions of the nucleon have been
renewed~\cite{Burkardt:2000za, Burkardt:2002hr, Miller:2007uy,
  Miller:2010nz, Lorce:2020onh, Jaffe:2020ebz}. Because of these
ambiguous relativistic corrections, the 3D charge and magnetization
distributions cannot be interpreted as quantum-mechanical probability
densities but can be understood as the unambiguous quasi-probabilistic
densities in the perspective of the Wigner
distributions~\cite{Wigner:1932eb, Hillery:1983ms, Lee:1995}. 

In the meanwhile, the generalized parton distributions (GPDs) of the
nucleon, which are defined in the infinite momentum frame (IMF) or on
the light cone, cast new light on its internal
structure~\cite{Mueller:1998fv, Ji:1996nm, Radyushkin:1997ki} (See also 
reviews~\cite{Goeke:2001tz, Diehl:2003ny, Ji:2004gf,
  Belitsky:2005qn}). Since Lorentz invariance ensures that the 
$(n+1)$th Mellin moments of a GPD should consist of an order-($n+1$)
polynomial, one can define the generalized form factors corresponding
to the matrix elements of the twist-2 and spin-($n+1$) local
operators. This indicates that the EM form factors of the nucleon can
be considered as the first ($n=0$) Mellin moment of the vector GPDs.   
The GPDs in the case of the purely transverse momentum transfer,
i.e. $\Delta^{+}=0 \, (\xi = 0)$, have naturally led to the
impact-parameter dependent parton  
distribution functions (PDFs)~\cite{Burkardt:2000za, Burkardt:2002hr},
which can be obtained by the two-dimensional (2D) Fourier transform of
the GPDs. Then, the first moment of the impact-parameter dependent
PDFs yields the 2D transverse charge distribution in the plane
perpendicular to the longitudinal momentum direction in the IMF. 
In contrast with the 3D charge and magnetization distributions in the
BF, the transverse charge distribution has a completely probabilistic
meaning. Miller~\cite{Miller:2007uy} computed the transverse charge
distributions of both the proton and neutron, based on the
experimental data on the EM form factors. In particular, the
transverse neutron charge distribution reveals a particularly
remarkable feature of the neutron structure: it becomes negative in
the core of the neutron. This is opposite to the 3D one, of which the
core part is positive, as was known for several
decades~\cite{Yennie:1957} and carved in many textbooks. Note that the
3D charge distribution is defined in the rest frame of the nucleon
whereas the 2D transverse one is presented in the IMF. Recently,
Lorc\'{e} described nicely how the BF and IMF distributions can be
naturally interpolated with each other~\cite{Lorce:2020onh}. When the
neutron is Lorentz-boosted, i.e., its longitudinal momentum approaches
infinity, the magnetic contribution dominates over the electric ones,
the central part of the transverse charge distributions turns to
negative. This means that in the BF where the longitudinal
momentum $P_z$ of the nucleon is equal to zero the charge distribution
is purely convective. When $P_z$ increases, a large contribution
induced by the magnetization drives the positive charges away from the
core part.  There is yet interesting and important aspect of the
transverse charge distributions. 

Very recently, Panteleeva and Polyakov~\cite{Panteleeva:2021iip} have
shown that how the BF distributions can be mapped directly onto
the IMF ones by using the Abel transformation~\cite{Abel} in the case of
the nucleon energy-momentum tensor. Thus, one can expect that the 
2D charge distributions in the IMF can be also directly
taken from the 3D ones in the BF with the help of the Abel
transformation. Considering the fact that the Abel transformation has
been used in quantum tomography~\cite{Leonhardt} as well as medical 
tomography~\cite{Natterer:2001}, one can regard this Abel image of the 
nucleon charge distribution as the nucleon tomography. Actually, the
Abel transformation has been introduced in the description of deeply
virtual Compton scattering~\cite{Polyakov:2007rv, Moiseeva:2008qd}.
In Ref.~\cite{Kim:2021jjf} the Abel images have been investigated for
the energy-momentum tensor distributions of the nucleon.  
When a particle has a spin higher than 1/2, the Radon
transformation~\cite{Radon, Panteleeva:2021iip} takes over the role of
the Abel transformation.  

In the present work, we first focus on the transverse charge
distributions of the nucleon when it is polarized along the
$x$-axis. So, the nucleon spin is aligned accordingly. As discussed  
already in Ref.~\cite{Carlson:2007xd}, the magnetic field starts to
induce an electric dipole field as the nucleon is Lorentz-boosted
along the $z$-direction, which is just a relativistic effect. We
will explicitly show how the strengths of both the proton and neutron
electric dipole moments get stronger as $P_z$ increases. These bring
about the distortion of the transverse charge distributions. We then
investigate how the Abel transformations map the 3D charge and 
magnetic distributions onto the 2D transverse ones in the IMF or in
the Drell-Yan frame (DYF). As concluded in
Refs.~\cite{Panteleeva:2021iip, Kim:2021jjf}, the 3D charge
distributions are equivalent to the corresponding 2D Abel images,
though the 2D ones have only complete probabilistic meaning. 

We sketch the present work as follows: In Sec.~\ref{sec:2}, we
present the formalism for the transverse charge
distributions in a moving polarized nucleon. Then we present the
numerical results for them as the longitudinal momentum varies. 
In Sect.~\ref{sec:3}, we show how to construct the 2D Abel images of
the 3D charge and magnetization distributions. The last Section is
devoted to the summary and conclusions of the present work.

%%%%%%%%%%%%%%%%%%%%%%%%%%%%%%%%%%%%%%%
\section{Transverse charge distributions in a moving nucleon
\label{sec:2}} 
%%%%%%%%%%%%%%%%%%%%%%%%%%%%%%%%%%%%%%%
\subsection{Formalism}
The 3D charge distribution of the nucleon cannot be
interpreted as a quantum-mechanical probabilistic density. Thus,
following Ref.~\cite{Lorce:2020onh}, we start with the
quantum-mechanical phase-space distribution, i.e., the   
Wigner distribution~\cite{Wigner:1932eb, Hillery:1983ms}. Since the
Wigner distribution is not positive-definite because of the
noncommutativity of the position and momentum operators, it cannot be
a probabilistic one. In the classical limit, it reduces to the
classical probabilistic distribution in phase space. 
This means that the 3D charge distribution can still be understood as
a quasi-probabilistic distribution, once it is defined through the
Wigner distribution. The nucleon matrix element of the EM current can
be defined as 
\begin{align}
\langle \hat{J}^{\mu}(\bm{r}) \rangle_{N}= \int
  \frac{d^{3}\bm{P}}{(2\pi)^{3}} d^{3}\bm{R} \, W_{N}(\bm{R},\bm{P})
  \langle \hat{J}^{\mu}(\bm{r}) \rangle_{\bm{R},\bm{P}}, 
\label{eq:1}
\end{align}
where $W_{N}(\bm{R},\bm{P})$ denotes the Wigner distribution that is
defined as 
\begin{align}
W_{N}(\bm{R},\bm{P}) &=\int \frac{d^{3}\bm{q}}{(2\pi)^{3}}
   e^{-i\bm{q}\cdot \bm{R}}   \tilde{\psi}_N^{*}\left(\bm{P} +
   \frac{\bm{\Delta}}{2}\right)   \tilde{\psi}_N\left(\bm{P} -
   \frac{\bm{\Delta}}{2}\right) \cr 
&=\int d^{3}\bm{z} \,e^{-i\bm{z}\cdot \bm{P}} {\psi}_N^{*}\left(\bm{R} -
  \frac{\bm{z}}{2}\right) {\psi}_N\left(\bm{R} +
  \frac{\bm{z}}{2}\right). 
\label{eq:Wig}
\end{align}
The average position $\bm{R}$ and momentum $\bm{P}$ are defined as
$\bm{R}=(\bm{r}_{f}+\bm{r}_{i})/2$ and
$\bm{P}=(\bm{p}_{f}+\bm{p}_{i})/2$, respectively. If one integrates
over the average position and momentum, then the probabilistic density
in either position or momentum space is found to be 
\begin{align}
  \int   \frac{d^{3}\bm{P}}{(2\pi)^{3}}\,W_{N}(\bm{R},\bm{P}) =|
  \tilde{\psi}_N\left(\bm{R} \right) |^{2},\;\;\;
\int d^{3}\bm{R}\,W_{N}(\bm{R},\bm{P}) =| {\psi}_N\left(\bm{P} \right)
  |^{2}.
\end{align}
The EM current operator is defined as 
\begin{align}
\hat{J}^{\mu}(0) = \bar{\Psi}(0) \gamma^\mu \hat{\mathcal{Q}} \Psi(0) 
\end{align}
with the quark field $\Psi(0)$ and the quark charge operator
$\hat{\mathcal{Q}}=\mathrm{diag}(2/3,\,-1/3,\,-1/3)$. 
The variable $\bm{z}=\bm{r}_f-\bm{r}_i$ represents the
position separation between the initial and final nucleons whereas
$\bm{\Delta}=\bm{p}_f-\bm{p}_i$ means the momentum transfer, which plays
a role of a probe for the structure of the nucleon. 

The Wigner distribution contains information on the wave packet of the 
nucleon 
\begin{align}
 \psi(\bm{r}) = \langle \bm{r} |\psi \rangle  = \int \frac{d^3
  \bm{p}}{(2\pi^3)} e^{i\bm{p}\cdot \bm{r}} 
  \tilde{\psi}(\bm{p}), \ \ \  \tilde{\psi}(\bm{p}) =
  \frac{1}{\sqrt{2p^{0}}}\langle p| \psi \rangle, 
\end{align}
with the plane-wave states $|\bm{p}\rangle$ and
  $|\bm{r}\rangle$ respectively normalized as  
$\langle \bm{p}_f|\bm{p}_i\rangle = 2 p_{i}^{0}(2\pi)^3
\delta^{(3)}(\bm{p}_f-\bm{p_i})$ and $\langle \bm{r}_f|\bm{r}_i\rangle
=  \delta^{(3)}(\bm{r}_f-\bm{r_i})$. The position state
$|\bm{r}\rangle$ is defined as a Fourier transform of the momentum
eigenstate $|\bm{p}\rangle$. If the nucleon were a 
nonrelativistic particle, then the momentum transfer would have only 
given rise to little change in the momentum of the nucleon. This means
that the nucleon would have been well localized. To put it more in
detail, only if the size of the wave packet $\delta |\bm{r}|$ is much
smaller than the size of the nucleon  $\delta |\bm{r}| \ll R_N$, it is
also much smaller than the resolution scale $\delta |\bm{r}| \ll
|\bm{\Delta}|^{-1}$, and it is much larger than the Compton wavelength
$\delta |\bm{r}| \gg 1/M_N$~\cite{Ji:2004gf, Belitsky:2005qn}, then
the nonrelativistic approximations would be validated for the nucleon
form factors, so that the nonrelativistic treatment suggested by 
Sachs~\cite{Ernst:1960zza,   Sachs:1962zzc} would have been valid for
the 3D charge distribution of the nucleon with the approximation
$\tilde{\psi}_N\left(\bm{P} \pm \bm{\Delta}/2\right) \approx
\tilde{\psi}_N(\bm{P})$. Unfortunately, the size of the nucleon is
comparable to the Compton wavelength, i.e., $R_N M_N\sim 1/4$, which
causes at least 20~\% of the relativistic effects on the 3D charge
distribution. Thus, Eq.~\eqref{eq:1}, which includes the
non-positive-definite Wigner distribution, retains a
quasi-probabilistic meaning. 

The  matrix element $\langle \hat{J}^{\mu}
(\bm{r})\rangle_{\bm{R},\bm{P}}$ conveys information on the internal
structure of the target localized around the average position $\bm{R}$
and average momentum $\bm{P}$. It can be expressed as the 3D Fourier 
transform of the matrix element $\langle  p_f, \lambda_f |
  \hat{J}^{\mu}(0) | p_i, \lambda_i \rangle$:  
\begin{align}
\langle \hat{J}^{\mu}(\bm{r}) \rangle_{\bm{R},\bm{P}}= \langle
  \hat{J}^{\mu}(0) \rangle_{-\bm{x},\bm{P}} = \int
  \frac{d^{3}\bm{\Delta}}{(2\pi)^{3}} e^{-i\bm{x} \cdot \bm{\Delta} }
  \frac{1}{\sqrt{2p^{0}_{i}}\sqrt{2p^{0}_{f}}}\langle  p_f, \lambda_f |
  \hat{J}^{\mu}(0) | p_i, \lambda_i \rangle, 
\end{align}
given at $x_{0}=0$ with the shifted position vector
$\bm{x}=\bm{r}-\bm{R}$. $\lambda_i(\lambda_f)$ denotes the helicity
of the initial (final) state of the nucleon. Note that both the
initial and final momentum satisfy the on-mass-shell conditions
$p^{2}_{i}=p^{2}_{f}=M_N^{2}$.  The matrix element of the EM current
operator  is parametrized in terms of the Dirac and Pauli form
factors:  
\begin{align}
\langle  p_{f}, \lambda_f | \hat{J}^{\mu}(0) | p_{i}, \lambda_i
  \rangle=  \overline{u}_{\lambda_f}(p_{f}) \left[\gamma^{\mu} F_1(t) +
  \frac{i\sigma^{\mu\nu} \Delta_{\nu}}{2m} F_2(t)\right]
  u_{\lambda_i}(p_{i}). 
\end{align}
The normalization of Dirac spinors is taken to be
$\overline{u}_{\lambda_f}(p)u_{\lambda_i}(p) = 2 M_N
\delta_{\lambda_f\lambda_i}$. As mentioned previously, we employ the
covariant normalization $\langle p_{f}, \lambda_f| p_{i}, \lambda_i
\rangle = 2p_i^{0} (2\pi)^{3}\delta_{\lambda_f  \lambda_i}
\delta^{(3)}(\bm{p}_{f}-\bm{p}_{i})$ for the one-particle states, and  
introduce the timelike average four-momentum
$P=(p_{i}+p_{f})/2$ and the spacelike four-momentum
transfer $\Delta=p_{f}-p_{i}$ with $\Delta^{2} = t$ such that $P$ and
$\Delta$ are orthogonal each other, $P\cdot \Delta =0$. 
Note that the spacelike momentum transfer $\bm{\Delta}$ lies in the
transverse plane. The frame that satisfy this condition, $\bm{P}\neq
\bm{0}$, and $\Delta^0 = 0$ is called the elastic frame
(EF)~\cite{Lorce:2017wkb, 
  Lorce:2018egm, Lorce:2020onh}. When $\bm{P}$ is taken to be a null
vector, it 
reduces to the BF in which the nucleon at rest is localized around
$\bm{R}$.  The matrix element of the EM current can be equivalently
parametrized in terms of the Sachs form factors, i.e., the electric
$G_{E}(t)$ and magnetic $G_{M}(t)$ form factors, which are expressed
as a linear combination of the Dirac and Pauli form factors  
\begin{align}
G_{E}(t) = F_{1}(t) - \tau F_{2}(t), \ \ \ G_{M}(t) = F_{1}(t) +
  F_{2}(t) 
\end{align}
with $\tau=  \bm{\Delta}^{2}/4M_N^{2}$.

We now define the 2D transverse EM distributions within the EF. 
The EF distributions depend on the impact parameter $b$
($\bm{x}=(\bm{b}, \, x_{z})$) and momentum $P_{z}$ where the nucleon   
moving along the $z$-direction without loss of generality: 
\begin{align}
 J^{0}_{\mathrm{EF}}(\bm{b},\,P_{z},\,\lambda',
  \,\lambda)= \int dx_{z} \langle
  \hat{J}^{0}(0) \rangle_{-\bm{x},\bm{P}}& = \int
 \frac{d^{2}\bm{\Delta}}{(2\pi)^{2}}
e^{-i \bm{b}\cdot \bm{\Delta}}
\frac{\langle p' \lambda'| \hat{J}^{0}(0) | p, \lambda
\rangle}{2P^{0}}\bigg{|}_{\Delta_{z}=0} .
\label{eq:j0}
\end{align}
When the nucleon is moving in the $z$-direction, the dependence on the
impact parameter and the elastic condition $\Delta^{0}=0$ are ensured
by integrating out the $x_z$ in position space. In this Section, we
show the unambiguous relativistic quasi-distributions for the
transversely polarized nucleon with $P_{z}$ varied. The advantage of
using the EF lies in the fact that it allows one to trace the origin
of the distorted charge distribution in the IMF in the Wigner sense.   

The transverse charge distribution of the nucleon in the
EF can be obtained by calculating the
  matrix element of the EM current in the momentum state given in
Eq.~\eqref{eq:j0} 
\begin{align}
J^{0}_{\mathrm{EF}}(\bm{b},P_{z},\lambda',\lambda)=  \int
  \frac{d^{2}\bm{\Delta}}{(2\pi)^{2}} 
  e^{-i \bm{b}\cdot  \bm{\Delta}} \left[
  g_{\mathrm{ch}}({\bm{\Delta}},P_{z})
 \delta_{\lambda' \lambda}
 +g_M({\bm{\Delta}},P_{z})
\frac{i (\sigma_{\perp}
\times
\bm{\Delta})_{\lambda'\lambda}}{2M_N}
\right], 
\label{eq:unpol}
\end{align}
where
$g_{\mathrm{ch},\,M} = g_{\mathrm{ch},M}^{E} +  g_{\mathrm{ch},M}^{M}$
with the electric and magnetic contributions 
defined separately as 
\begin{align} 
&g_{\mathrm{ch}}^E (\bm{\Delta},P_{z})=
  \frac{P_{0}+M_N(1+\tau)}{(P_{0}+M_N)(1+\tau)} G_{E}(t) , \ \ \
  g_{\mathrm{ch}}^M (\bm{\Delta},P_{z})= \frac{\tau
  P^{2}_{z}}{P_{0}(P_{0}+M_N)(1+\tau)}G_{M}(t), \cr 
&g_M^{E}(\bm{\Delta},P_{z})=
  -\frac{P_{z}}{(P_{0}+M_N)(1+\tau)} G_{E}(t), \ \ \
  g_M^{M}(\bm{\Delta},P_{z})=
  \frac{P_{z}(P_{0}+M_N+\tau M_N)}{P_{0}(P_{0}+M_N)(1+\tau)} G_{M}(t) .
\label{eq:gem}
\end{align}
Here, $P^{0}$ is given by $\sqrt{(1+\tau)M_N^2+P^{2}_{z}}$.
The charge distribution of the unpolarized  nucleon is obtained by
taking the trace of $J_{\mathrm{EF}}^0$ in spin space, which is
solely due to $g_{\mathrm{ch}}$. Thus, $J_{\mathrm{EF}}^0$ can be
identified as $\rho_{\mathrm{ch}}(b,P_z)$. On the other hand, $g_{M}$
comes into play only when the nucleon is transversely polarized, which
brings about the deformation of the charge
distribution~\cite{Carlson:2007xd}. The EF charge (magnetization)
distribution receive the magnetization (charge) contribution when the
nucleon starts being boosted, since the temporal and spatial
components of the four current are mixed by the 
Lorentz boost.  When the nucleon is polarized along $x$-axis, i.e. its
spin state is taken to be $|s_x \rangle= (|\lambda =1/2 \rangle +
|\lambda =-1/2 \rangle)/\sqrt{2}$, the transverse charge distribution
of the transversely polarized nucleon in the EF can be written as  
\begin{align}
\rho_{\mathrm{ch}}^{T}(\bm{b},\,P_{z}) 
&=  \rho_{\mathrm{ch}}(b,P_{z}) -\frac{1}{2M_N}
  \frac{d}{db_y} \rho_{M}(b,P_{z}) 
\label{eq:transch}
\end{align}
with
\begin{align}
\rho^{E,M}_{\mathrm{ch,M}}(b,P_{z})=\int
\frac{d^{2}\bm{\Delta}}{(2\pi)^{2}}
 e^{-i\bm{b}\cdot
 \bm{\Delta}}
g^{E,M}_{\mathrm{ch,M}}(\bm{\Delta},P_{z}).    
\end{align}
Note that $\rho_{\mathrm{ch}}^T$ contains explicitly the contribution
of the magnetization, represented by the second term of
Eq.~\eqref{eq:transch}.  

One can now see that the charge distributions of the transversely
polarized nucleon in the EF exhibits how the 2D charge distributions
in the BF are interpolated to the those in the IMF. That is,
one can show that these charge distributions in 
the EF with $P_{z}\to \infty$ coincide with those in the
IMF. If we take the limit $P_{0} \to P_{z}=\infty$, which brings one
from the BF to the IMF,  the charge and magnetization parts of the
expressions given in Eq.~\eqref{eq:gem} are respectively reduced to
the Dirac and Pauli form factors: 
\begin{align} 
g_{\mathrm{ch}} (\bm{\Delta},P_{z}=\infty)=   F_{1}(t),\;\;\; 
g_M(\bm{\Delta},P_{z}=\infty)=
 F_{2}(t).
\end{align}
Then the transverse charge distribution of the unpolarized
nucleon in the IMF is obtained as the 2D Fourier transform of $F_1$ 
\begin{align}
\rho_{\mathrm{ch}}(b) =  \int
  \frac{d^{2}\bm{\Delta}}{(2\pi)^{2}} e^{-i \bm{b}\cdot   \bm{\Delta}}
  F_{1}(t),
\label{eq:FrF1}
\end{align}
and that of the transversely polarized nucleon in the IMF is written
as  
\begin{align}
\rho_{\mathrm{ch}}^{{\mathrm{T}}}(\bm{b}) = \int
  \frac{d^{2}\bm{\Delta}}{(2\pi)^{2}}   e^{-i \bm{b}\cdot  
\bm{\Delta}}   F_{1}(t)  +   i\int \frac{d^{2}\bm{\Delta}}{(2\pi)^{2}}
 e^{-i \bm{b}\cdot \bm{\Delta}} \frac{\Delta_{y}}{2M_N} F_{2}(t) .
\end{align}
On the other hand, if we assume that the nucleon is at rest ($P_z=0$),
we can easily show that the charge and magnetization distributions in
the EF coincide with those in the BF:   
\begin{align}
&\int dx_{z}
J^{0}_\mathrm{EF}(\bm{x},P_{z}=0)= \int
  \frac{d^{2}\bm{\Delta}}{(2\pi)^{2}} e^{-i\bm{b}\cdot 
\bm{\Delta}}\frac{m}{P^{0}}  G_{{E}}(t) \delta_{\lambda',\lambda},  \cr 
& \int dx_{z} J^{i}_{\mathrm{EF}}(\bm{x},P_{z}=0)= 
\int \frac{d^{2}\bm{\Delta}}{(2\pi)^{2}}
 e^{-i \bm{b}\cdot \bm{\Delta}} \frac{1}{2P^{0}} i\varepsilon^{ki}
\Delta_{k}\sigma^{3}_{\lambda' \lambda}G_{{M}}(t).
\end{align}
When the nucleon is not boosted ($P_{z}=0$), i.e., when it is in the
rest frame, there is no distortion of the charge distribution of the
transversely polarized nucleon. However, if we increase the 
longitudinal momentum of the nucleon $P_z$, the charge distribution
starts to be distorted. This deformation arises from the induced
electric dipole moment, which appears naturally by Einstein's special
theory of relativity~\cite{Carlson:2007xd}. The
induced electric dipole moment is defined as  
\begin{align}
&d_{y}(P_{z}) := \frac{e}{2M_N} \int d^{2}\bm{b} \,
  \rho_{\mathrm{M}}(b, P_{z}),   
\label{eq:elecdip}
\end{align}
which shows how the charge distribution is deformed from the
spherically symmetric shape quantitatively. The electric dipole moment
in the IMF is found to be $d^{p,n}_{y}=e\kappa_{p,n} /2M_N$. The
corresponding expression is indeed restored from the EF
$d^{p,n}_{y}(P_{z}=\infty)=e\kappa_{p,n}/2M_N$ by boosting the
nucleon. Interestingly, the charge distribution is maximally distorted
at $P_{z}=3.2$~GeV. 

\subsection{Results and discussion}
\begin{figure}[htp]
\includegraphics[scale=0.285]{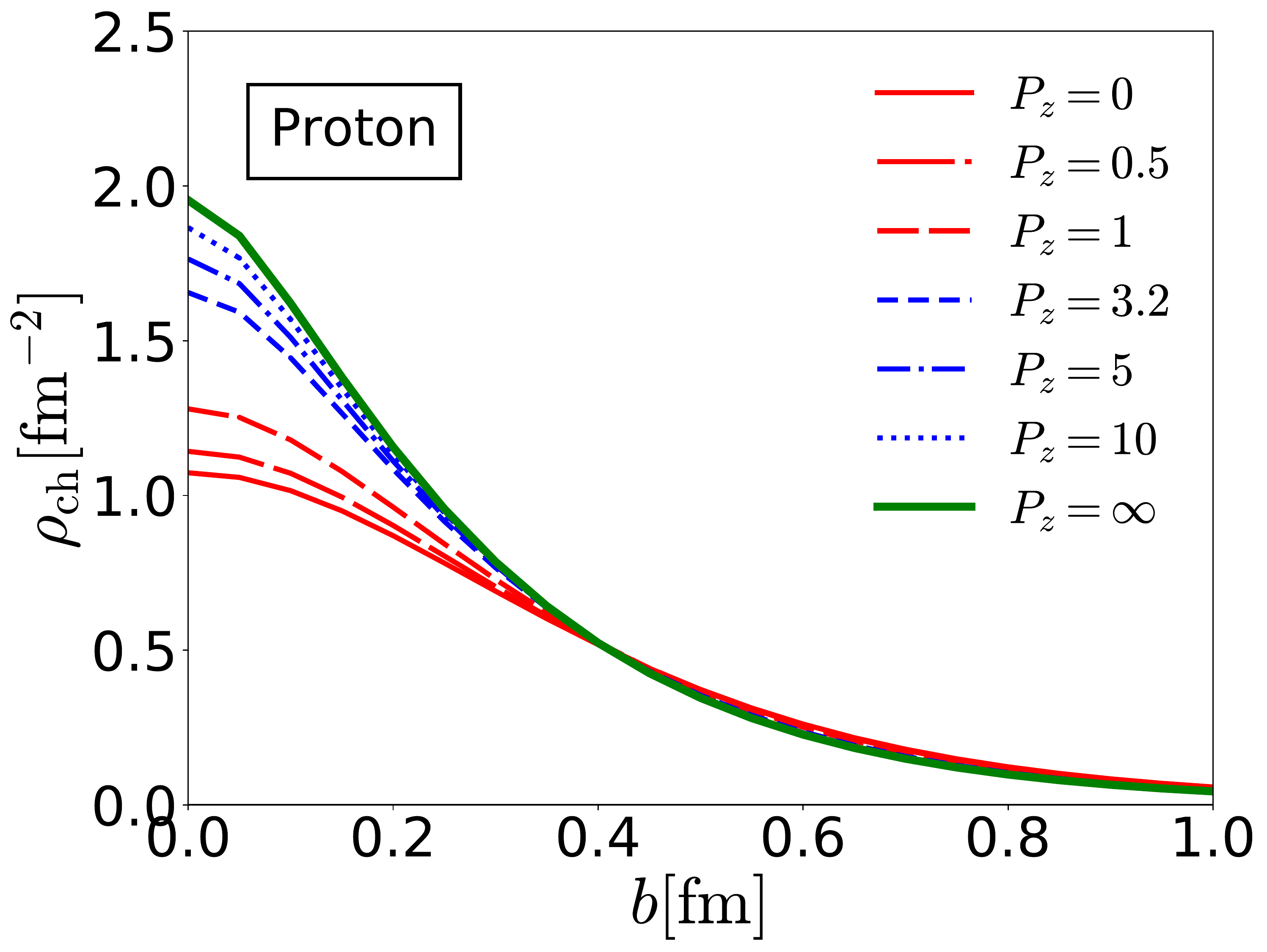}
\includegraphics[scale=0.285]{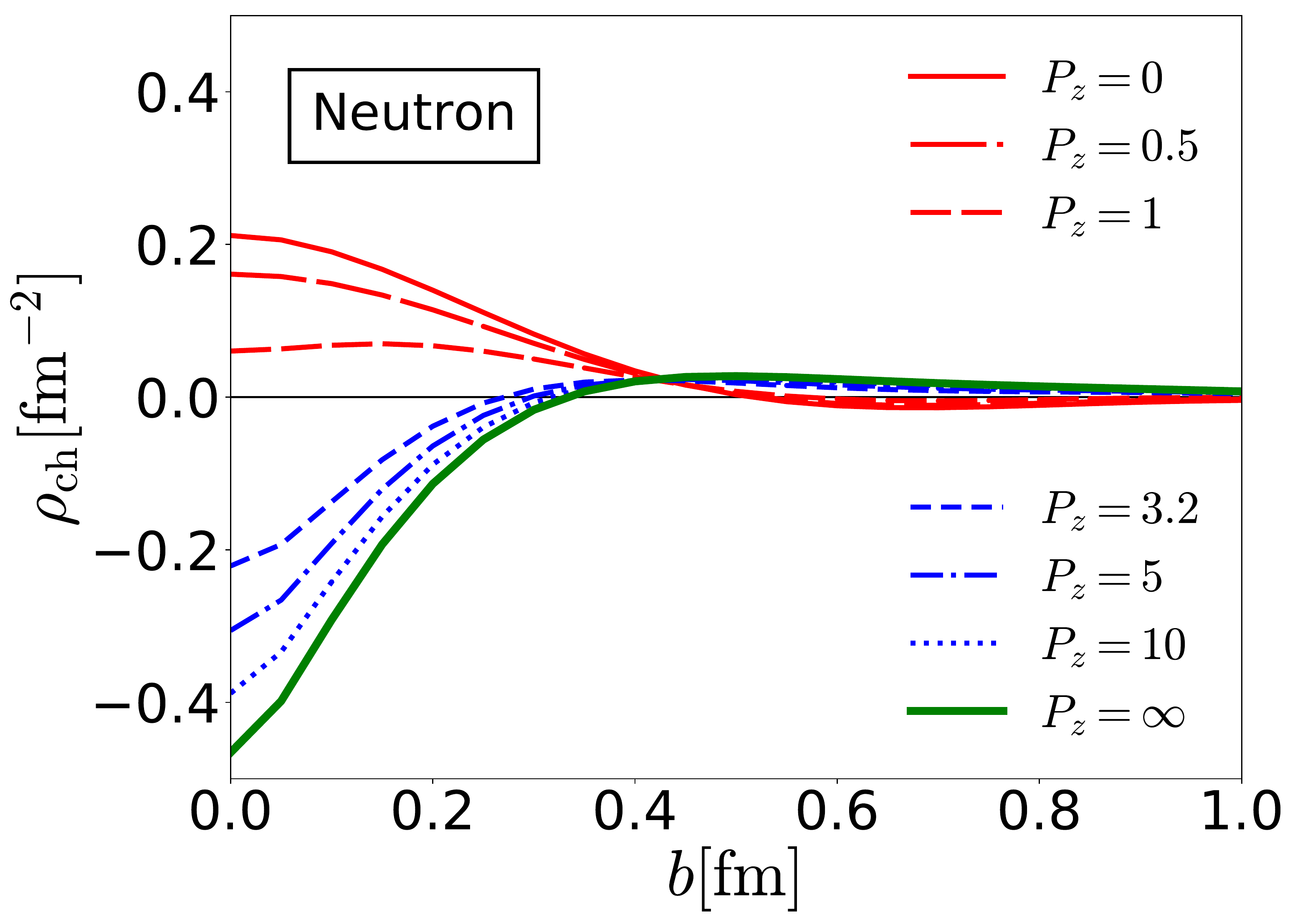}
\caption{The left (right) panel depicts the transverse charge
  distributions of the unpolarized proton (neutron) with $P_{z}$ varied
  from zero to infinity. The EM form factors of both the proton and
  neutron are taken from the parametrizations given in
  Ref.~\cite{Bradford:2006yz}. }  
\label{fig:1}
\end{figure}
\begin{figure}[htp]
\includegraphics[scale=0.285]{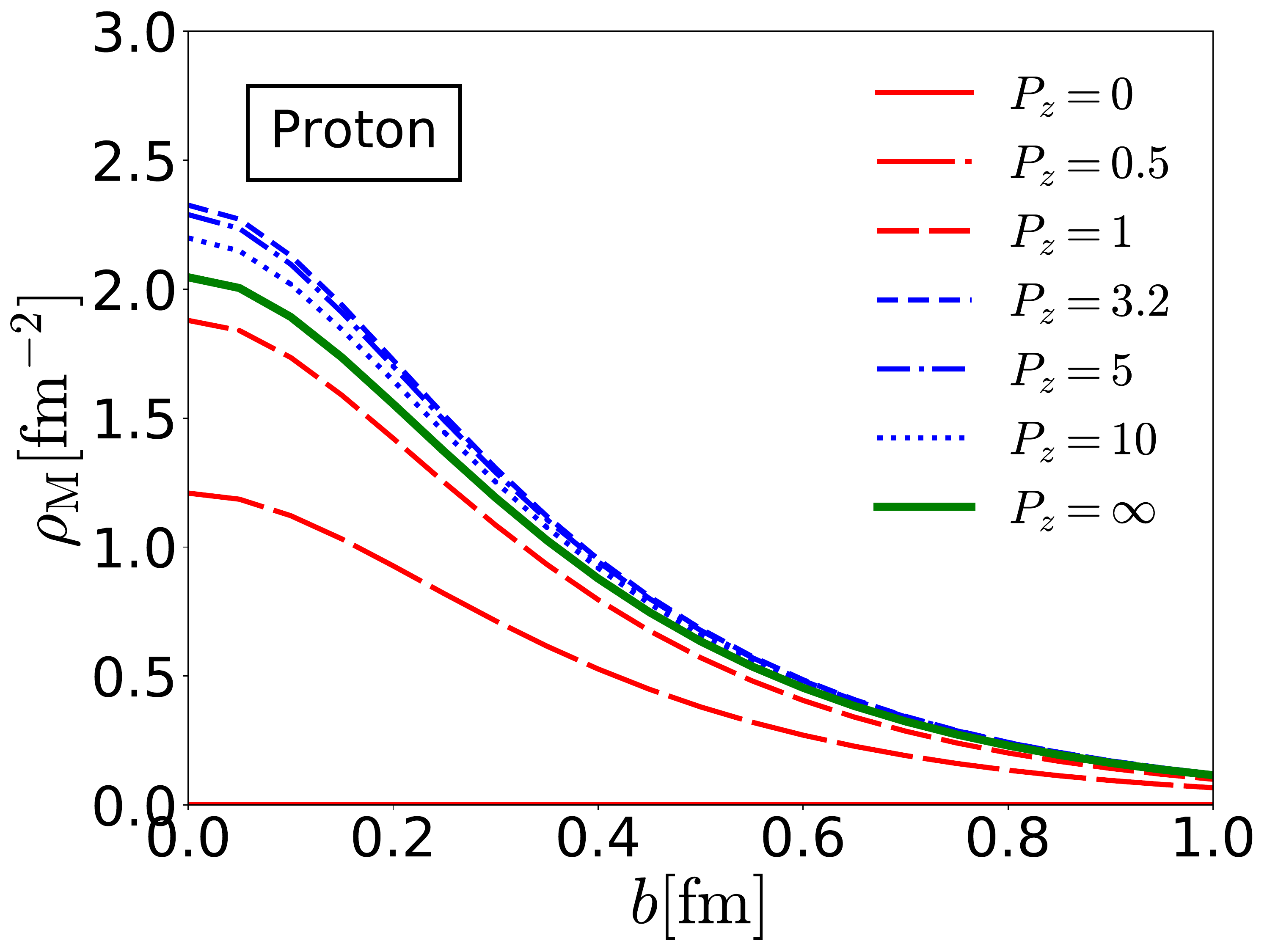}
\includegraphics[scale=0.285]{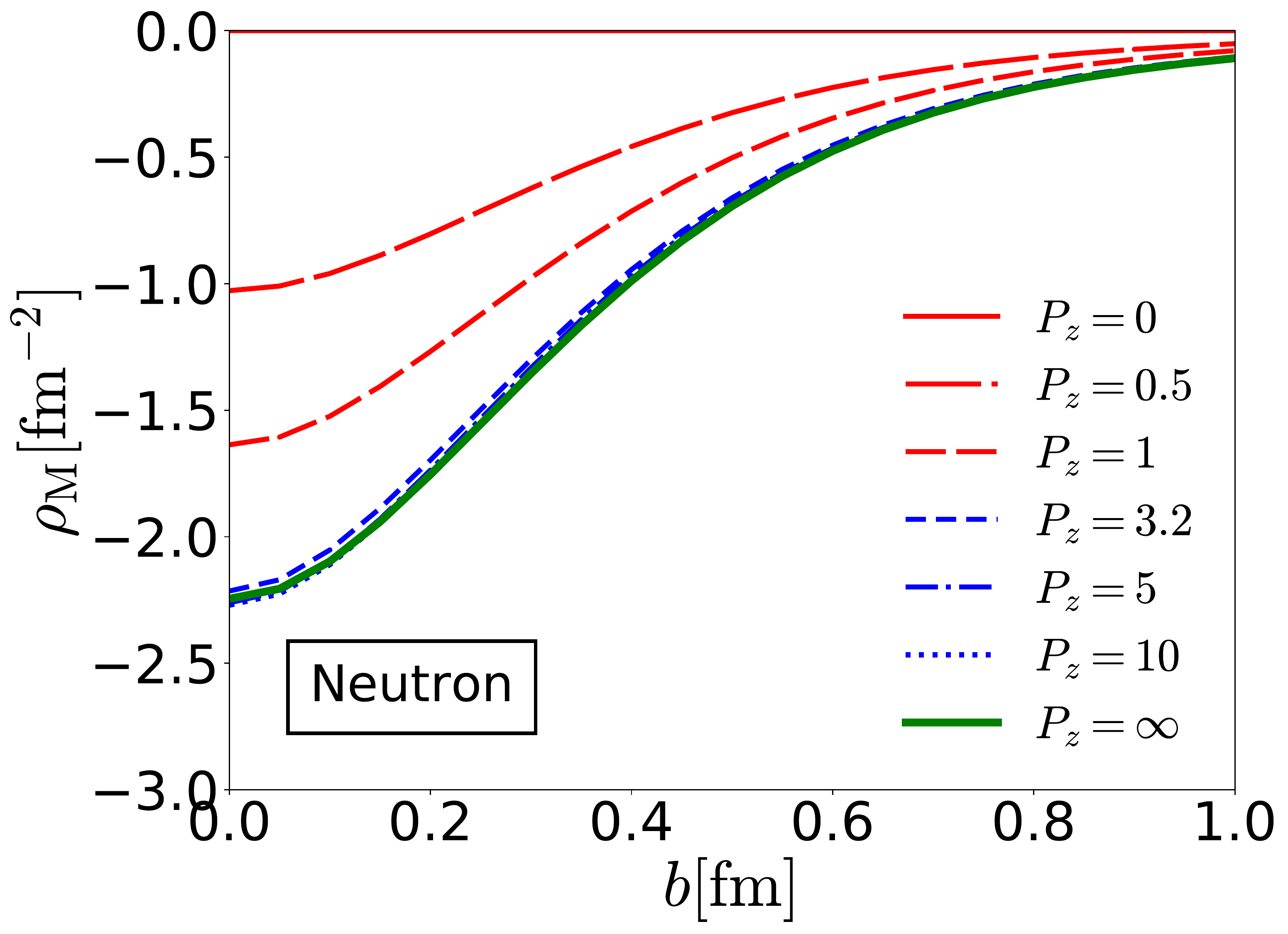}
\caption{The left (right) panel depicts the transverse magnetization 
  distributions of the unpolarized proton (neutron) with $P_{z}$ varied
  from zero to infinity. The EM form factors of both the proton and
  neutron are taken from the parametrizations given in
  Ref.~\cite{Bradford:2006yz}.} 
\label{fig:2}
\end{figure}
We are now in a position to present and discuss the numerical results
of the charge distributions of the unpolarized and transversely
polarized nucleons. To derive them, we employ the empirical EM form
factors of the nucleon extracted from the experimental
data~\cite{Bradford:2006yz}. We first show how the charge
distribution of the unpolarized proton undergoes changes as $P_z$
increases, which has already been studied in
Ref.~\cite{Lorce:2020onh}. For completeness, we present the
corresponding results in Fig.~\ref{fig:1}. As explained in
Ref.~\cite{Lorce:2020onh} in detail, the core part of the charge
distribution of the proton gets larger as $P_z$ increases from $P_z=0$
to $P_z=\infty$ monotonically whereas the tail part slowly gets lessened. 
As for the neutron, the core part decreases and turns negative as
$P_z$ increases, while the tail part gets strengthened and turns
positive. This explains the apparent difference between the transverse
charge distribution of the neutron in the IMF and the 3D
one~\cite{Miller:2007uy, Miller:2010nz}. In Fig.~\ref{fig:2}, we show
how the magnetization distributions of the proton and neutron undergo
changes as $P_z$ increases. The general behavior of the magnetization
distributions seems to be similar to that of the charge ones. However,
as we can see from the left panel of Fig.~\ref{fig:2}, the
magnetization distribution of the proton is maximized at $P_z\approx
3.2$ GeV, not at $P_z=\infty$. It can be originated from the
kinematical factors in Eq.~\eqref{eq:gem}. So, this effect is not
dynamical but kinematical as shown from Eqs.~\eqref{eq:unpol} and
\eqref{eq:gem}.  

\begin{figure}[htp]
\centering
\includegraphics[scale=0.285]{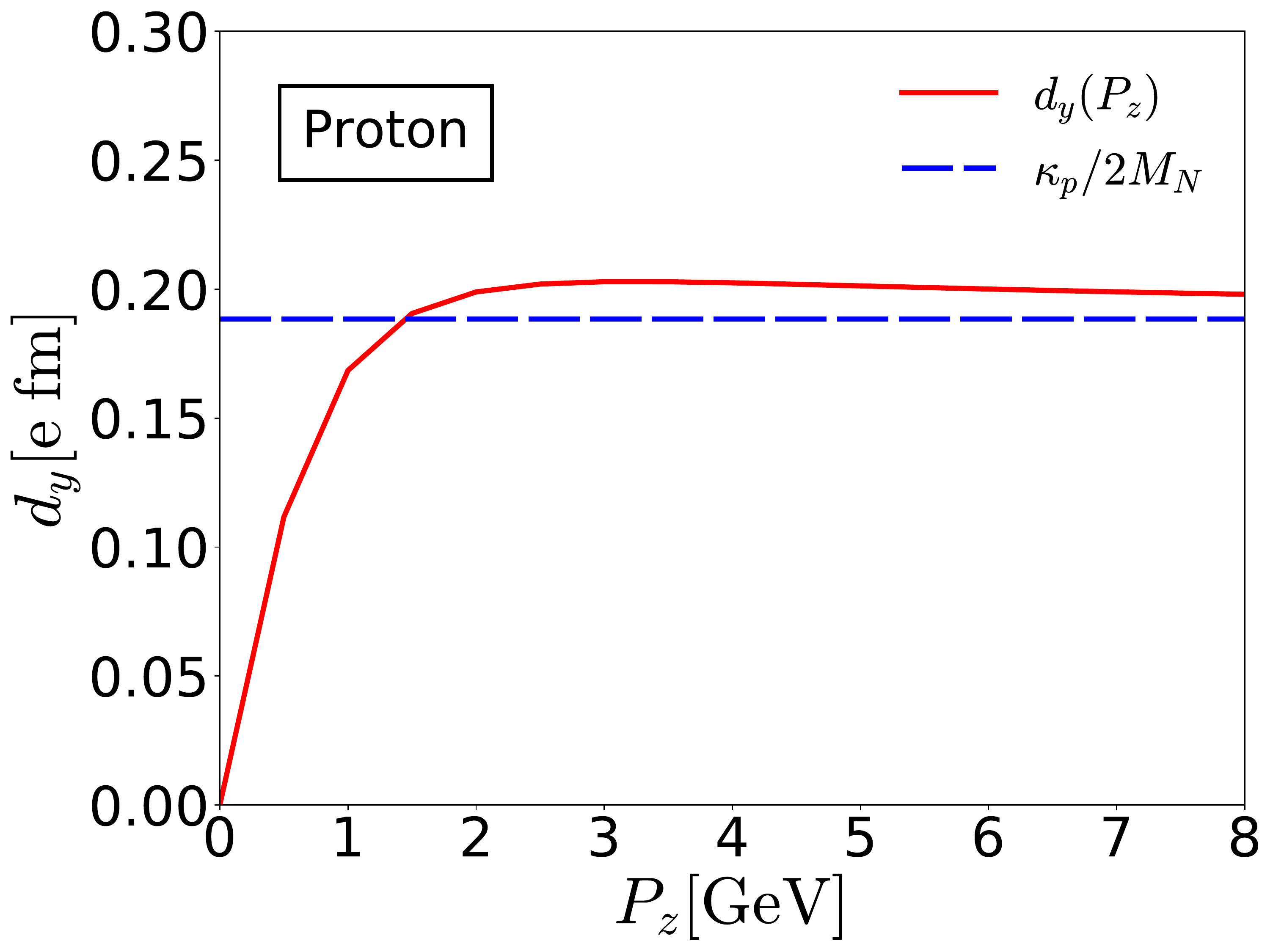}
\includegraphics[scale=0.285]{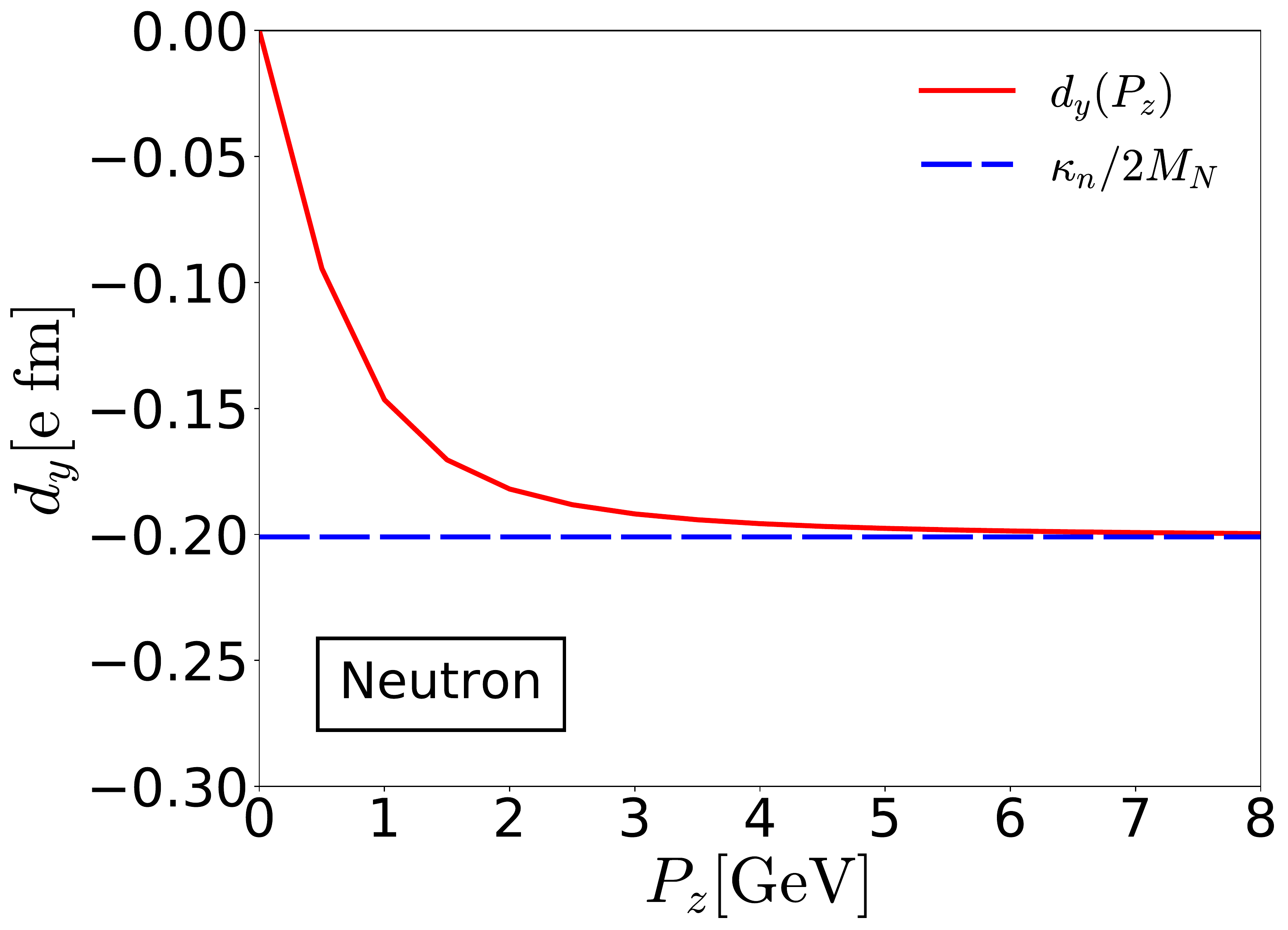}
\caption{The left (right) panel depicts the induced electric dipole
  moment of the proton (neutron) as a function of $P_{z}$. The dashed
  blue line draws the value of the induced electric dipole moment of
  the nucleon in the IMF.} 
\label{fig:3}
\end{figure}
\begin{figure}[htp]
\centering
\includegraphics[scale=0.300]{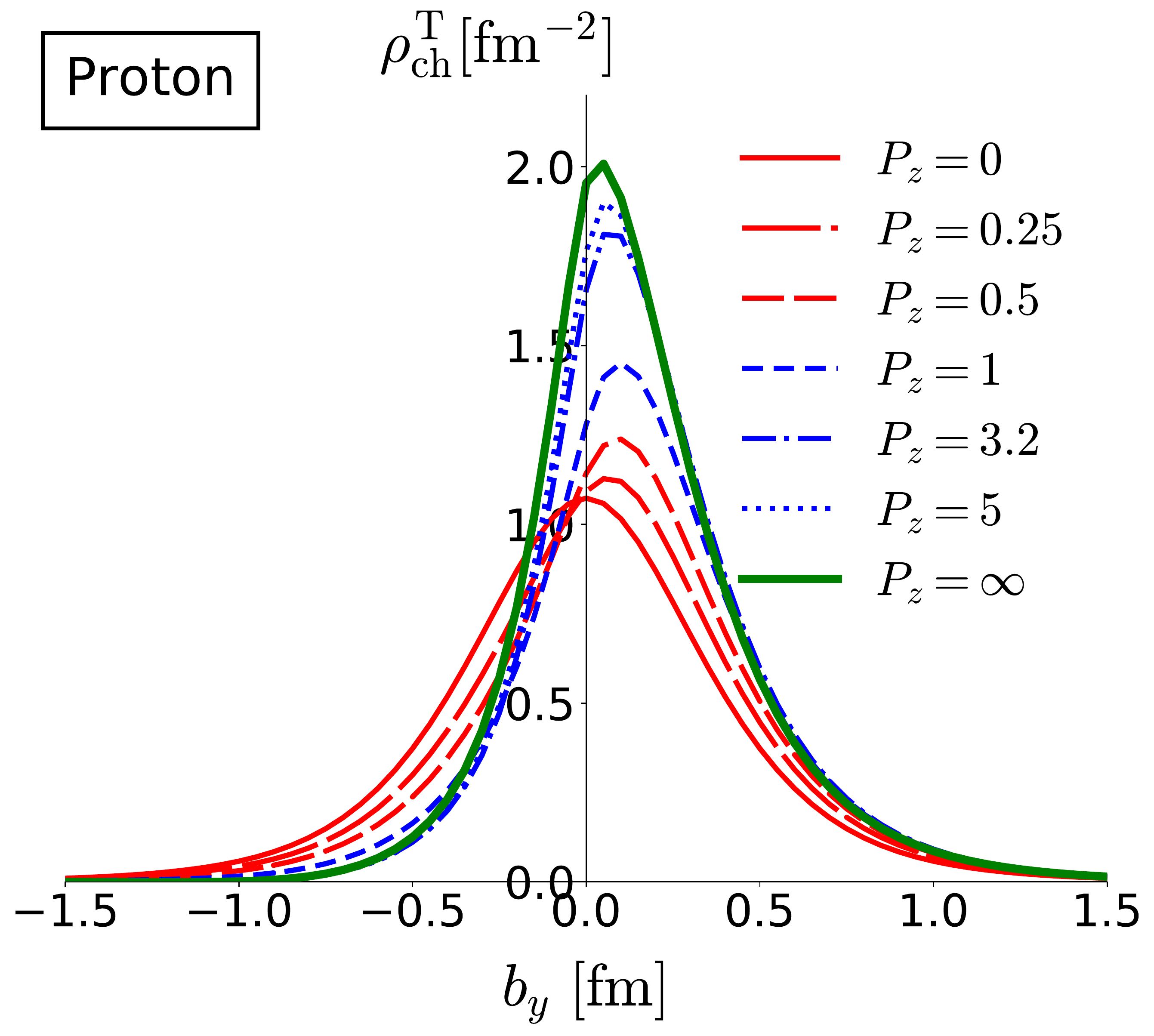}
\includegraphics[scale=0.300]{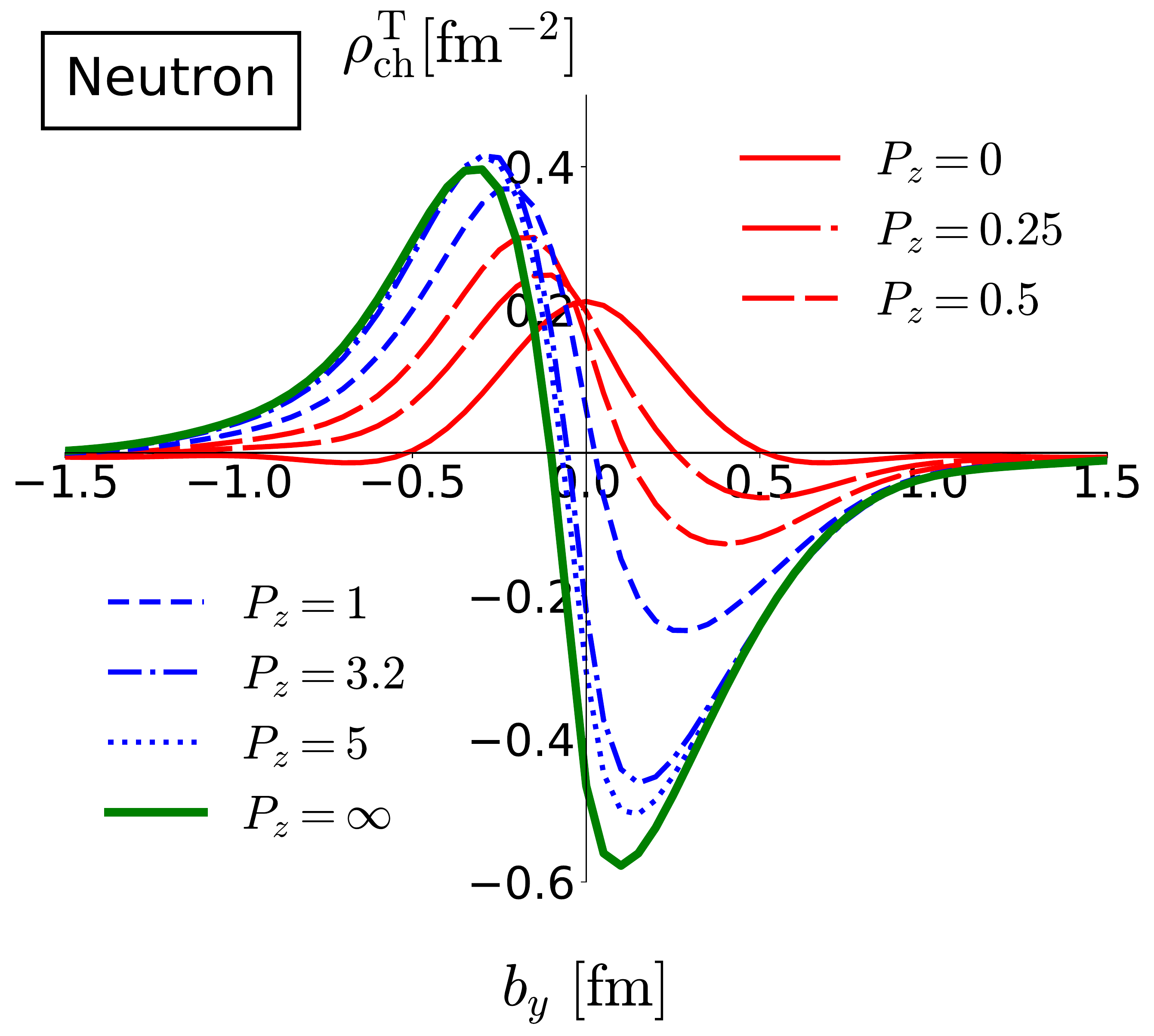}
\caption{The left (right) panel depicts the transverse charge
  distributions of the transversely polarized proton (neutron) along
  the $b_x$-direction, with $P_{z}$ varied from zero to infinity. The
  EM form factors of both the proton and neutron are taken from the
  parametrizations given in Ref.~\cite{Bradford:2006yz}. } 
\label{fig:4} 
\end{figure}
\begin{figure}[htp]
\includegraphics[scale=0.39]{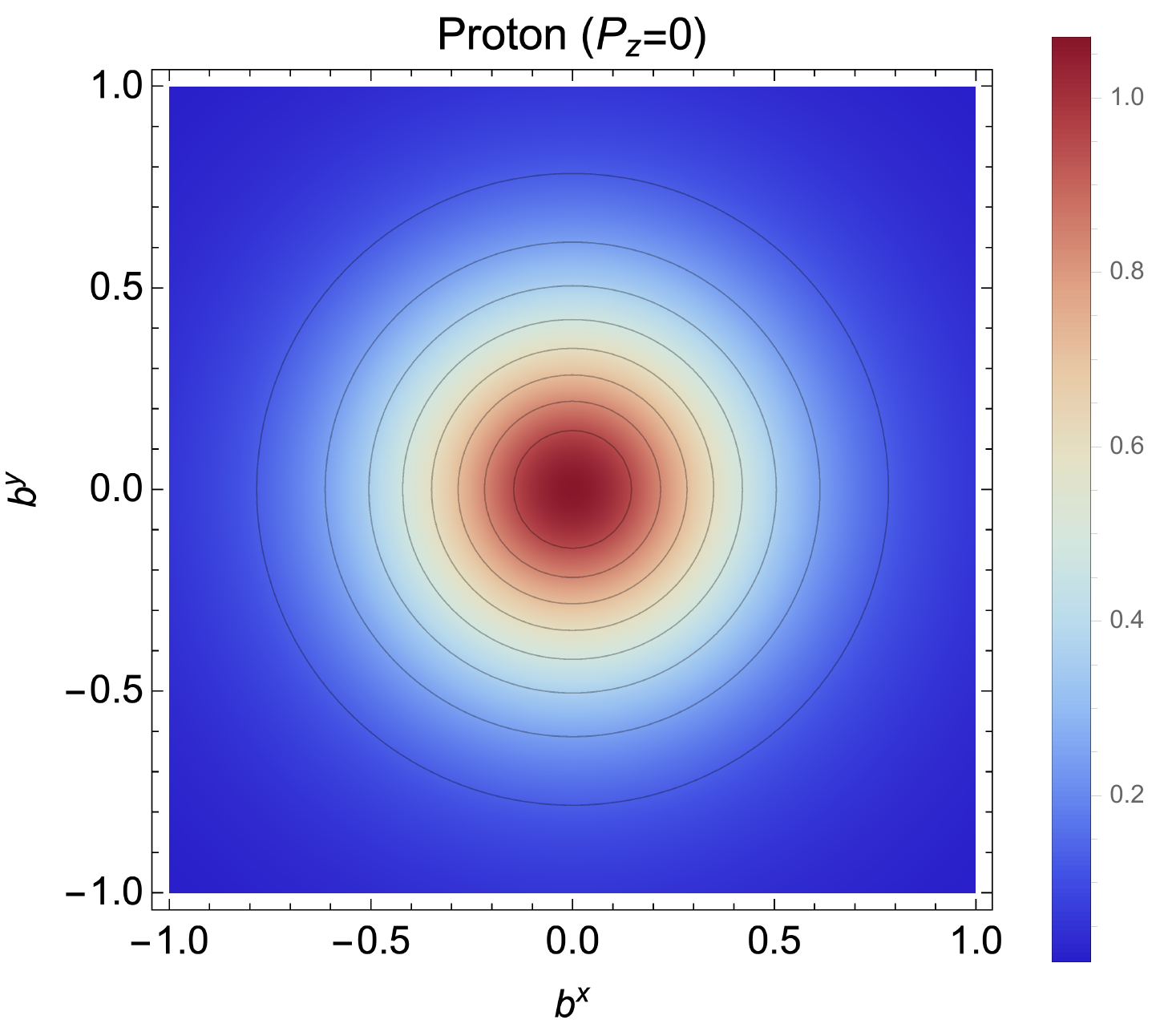}
\includegraphics[scale=0.39]{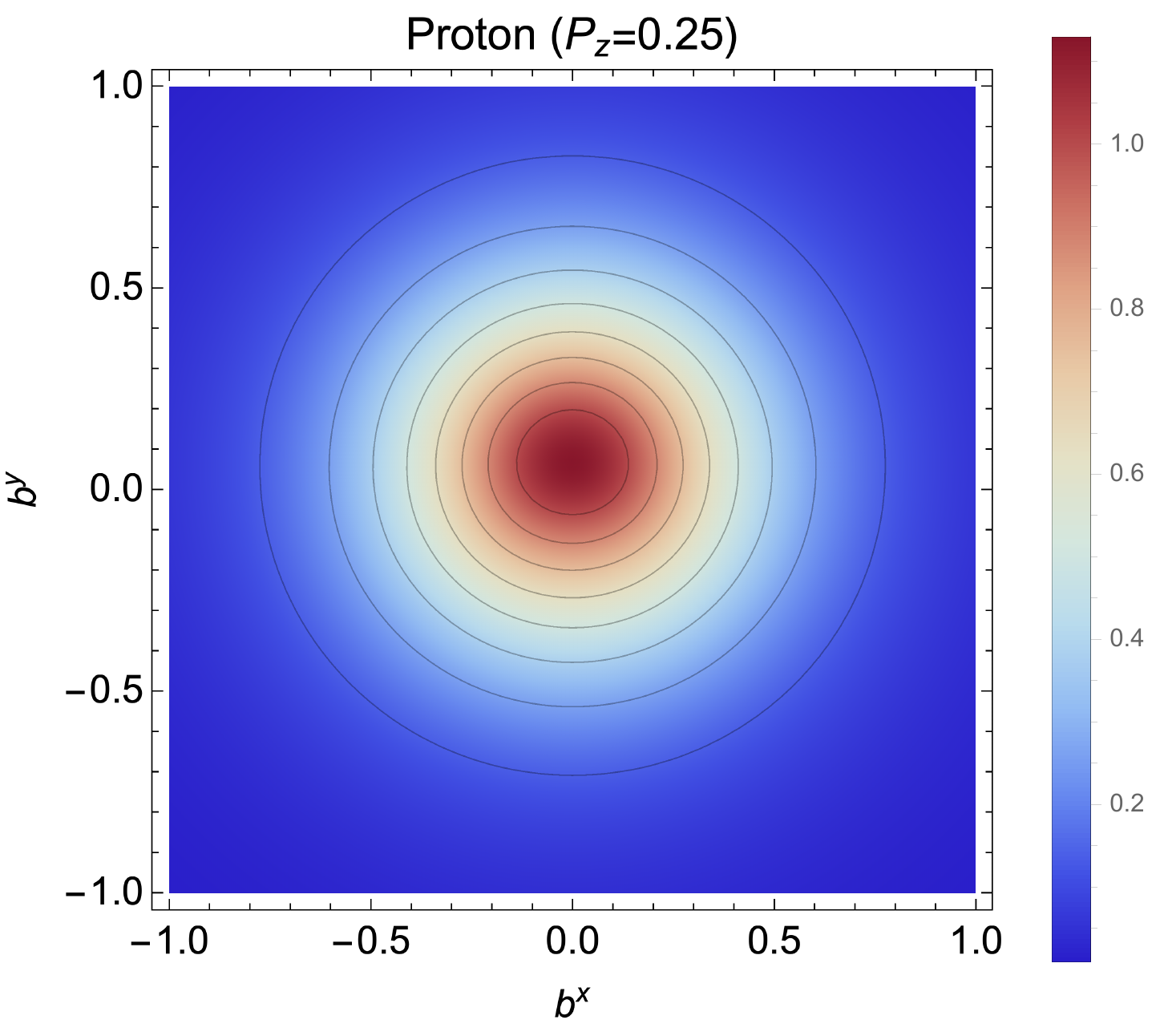}
\includegraphics[scale=0.39]{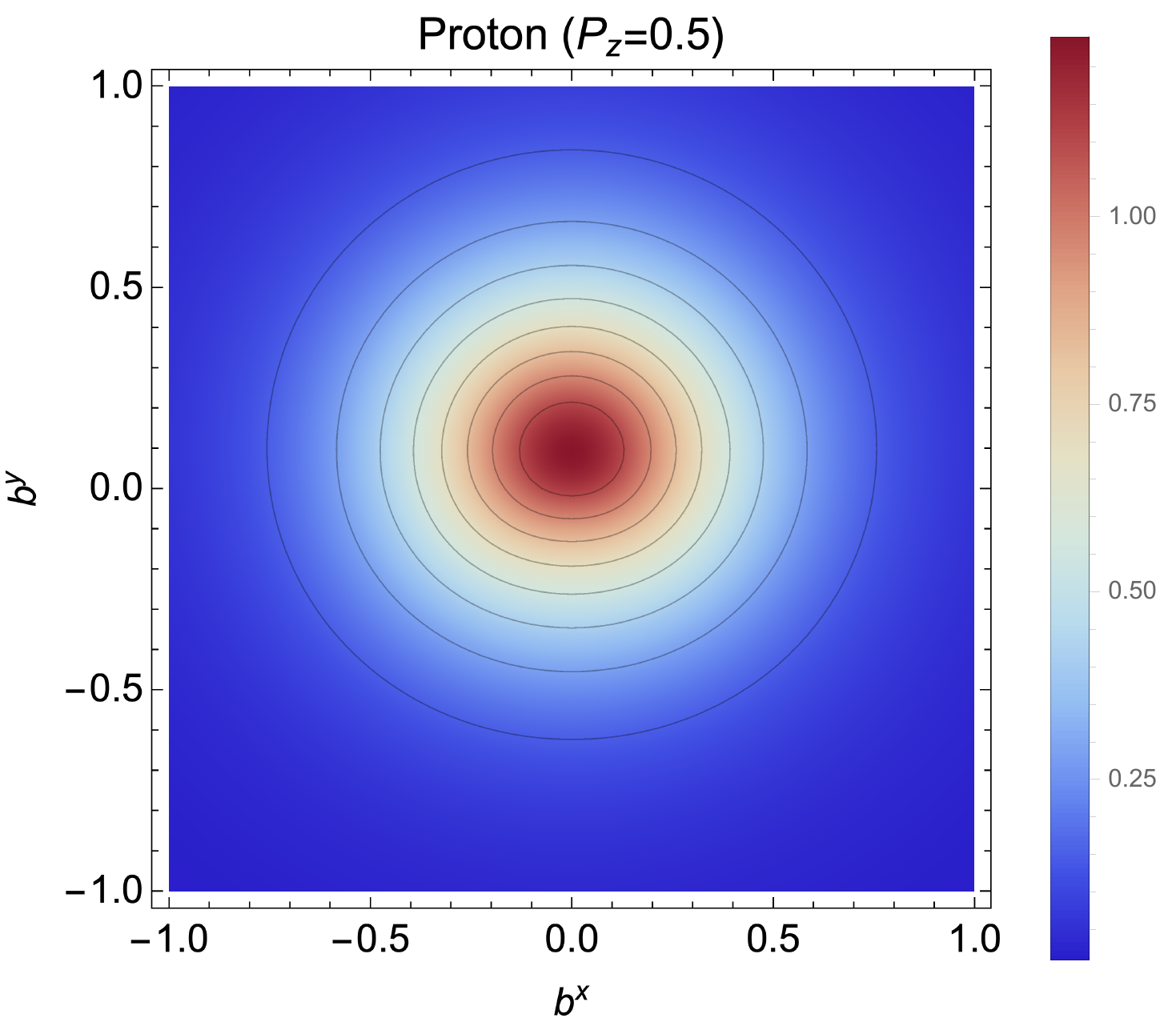}
\includegraphics[scale=0.39]{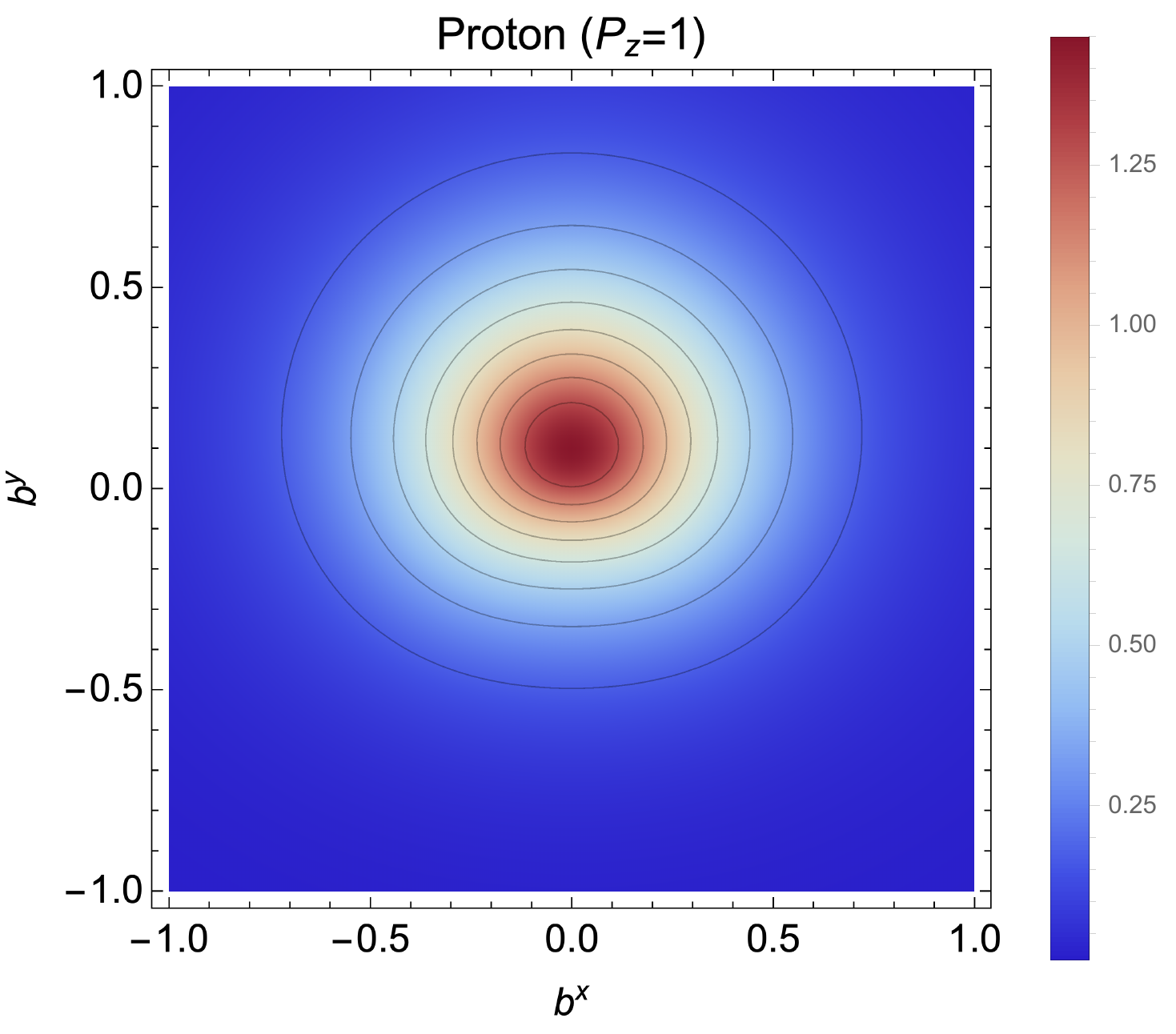}
\includegraphics[scale=0.39]{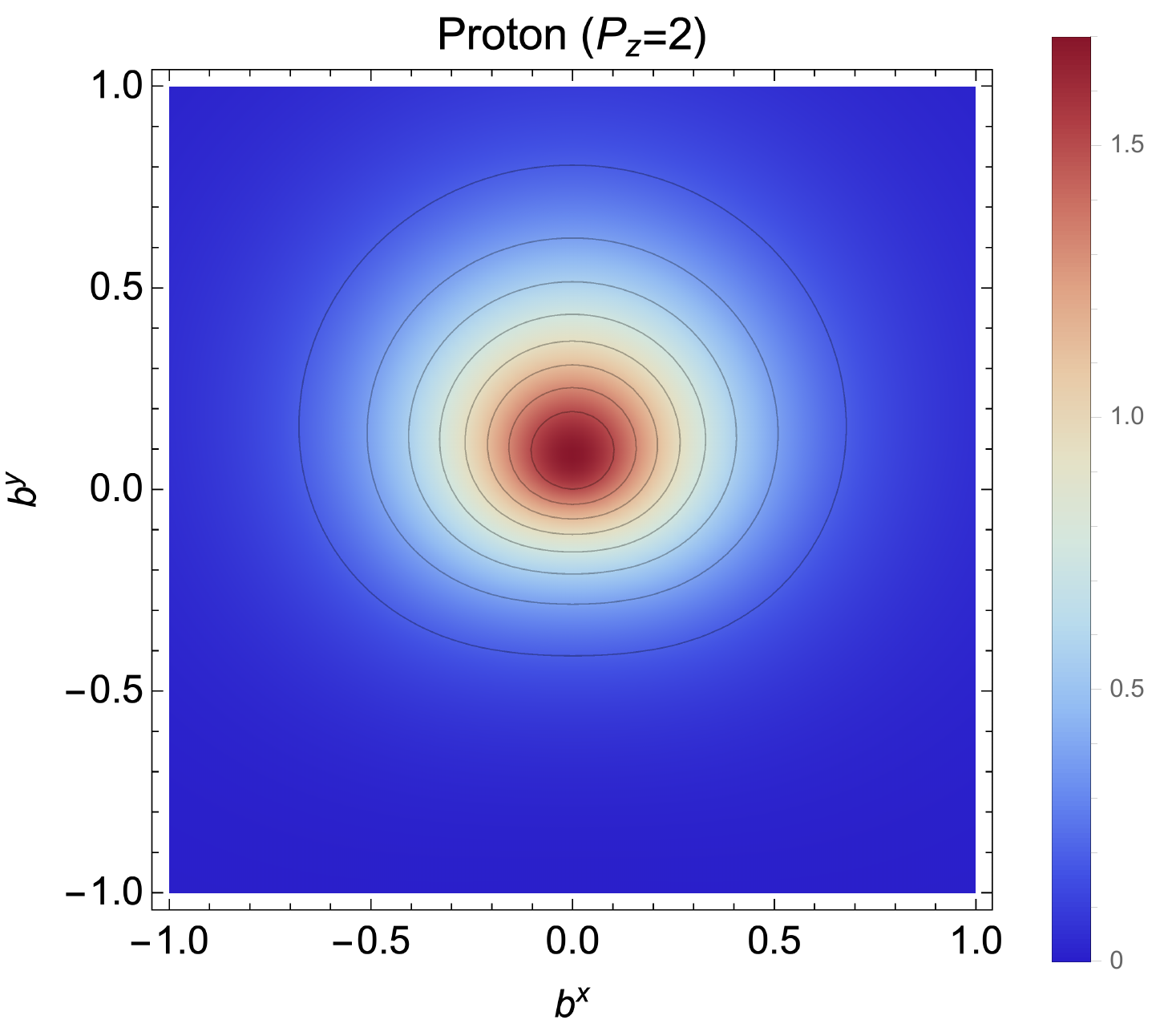}
\includegraphics[scale=0.39]{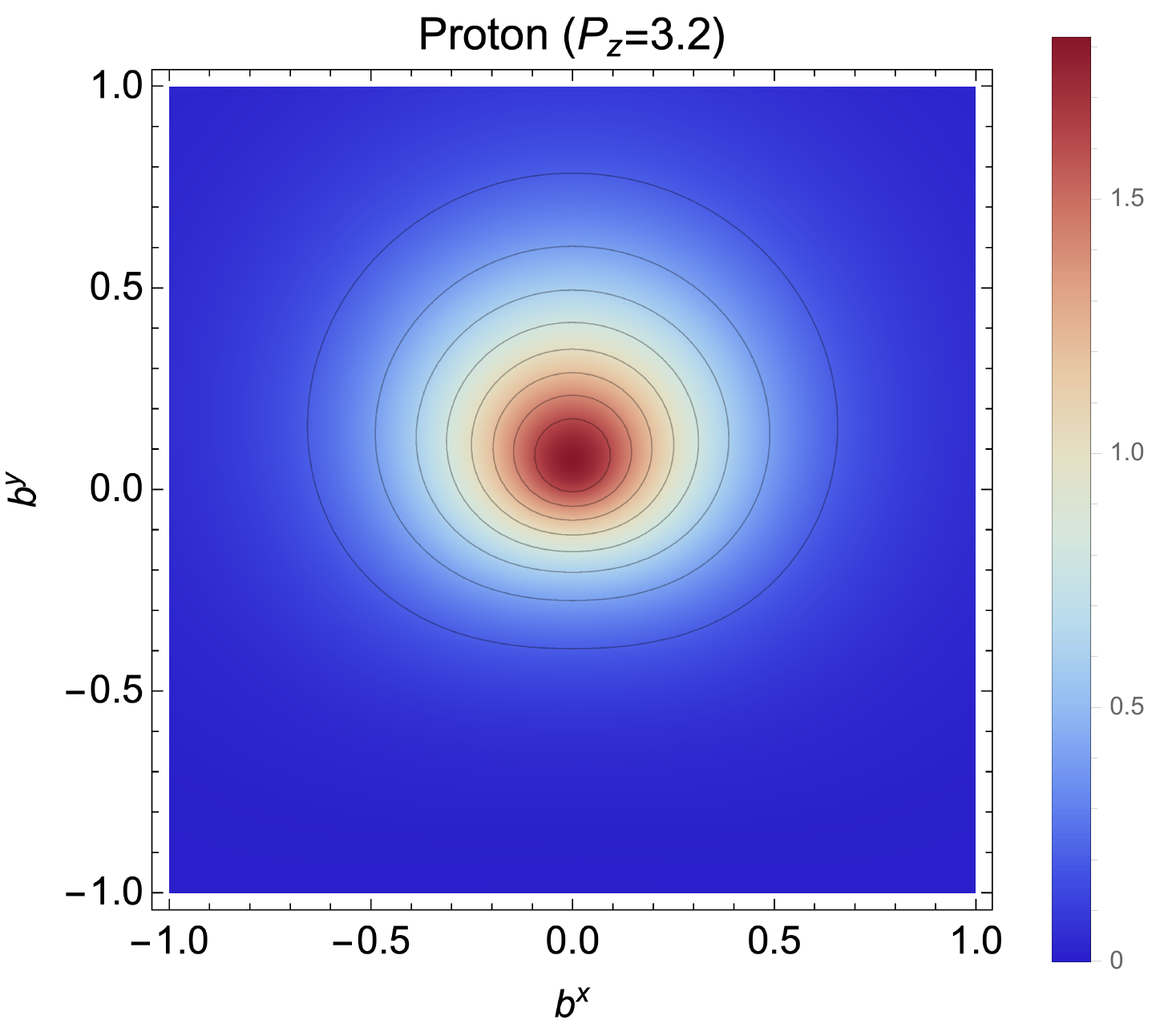}
\includegraphics[scale=0.39]{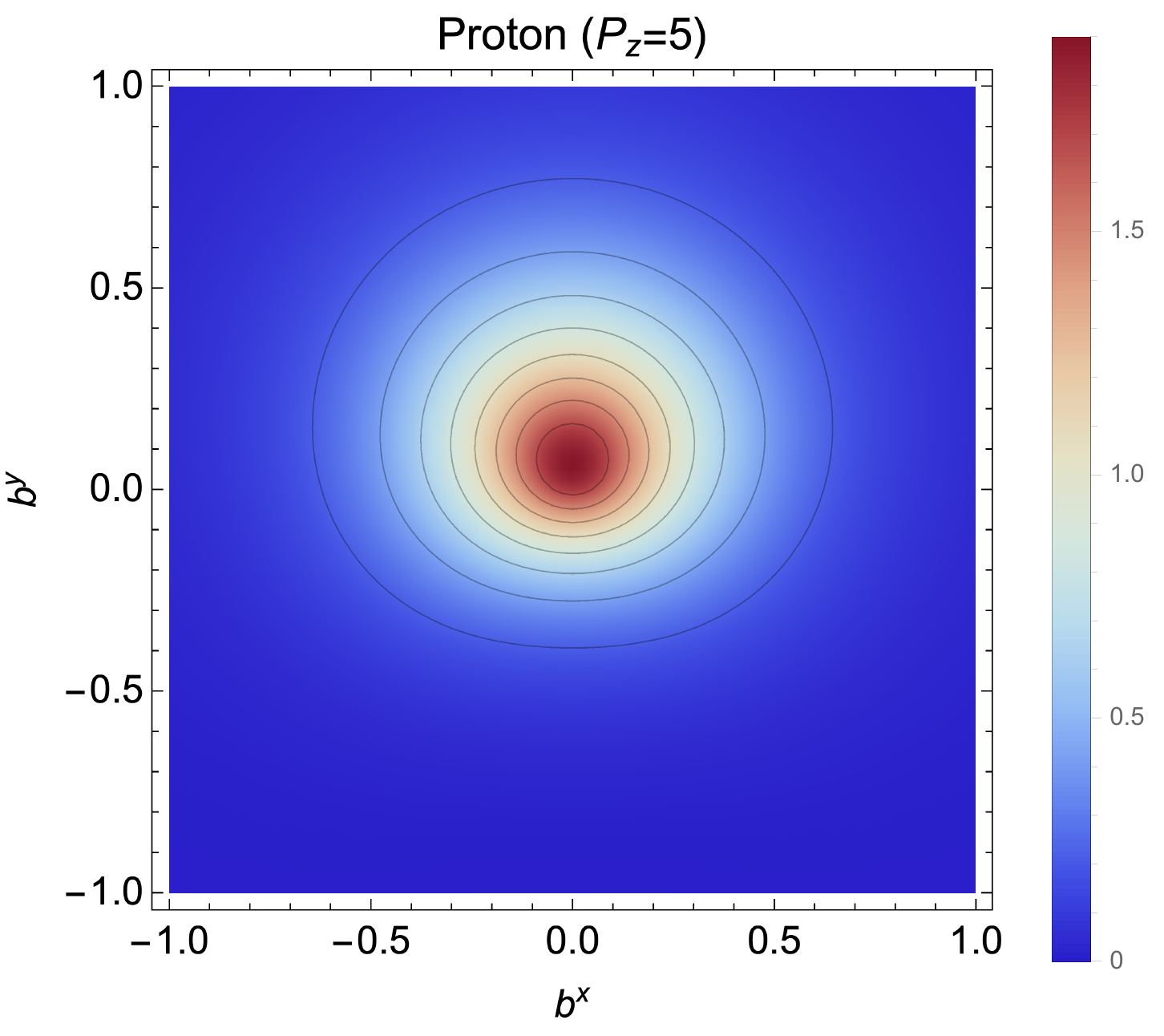}
\includegraphics[scale=0.39]{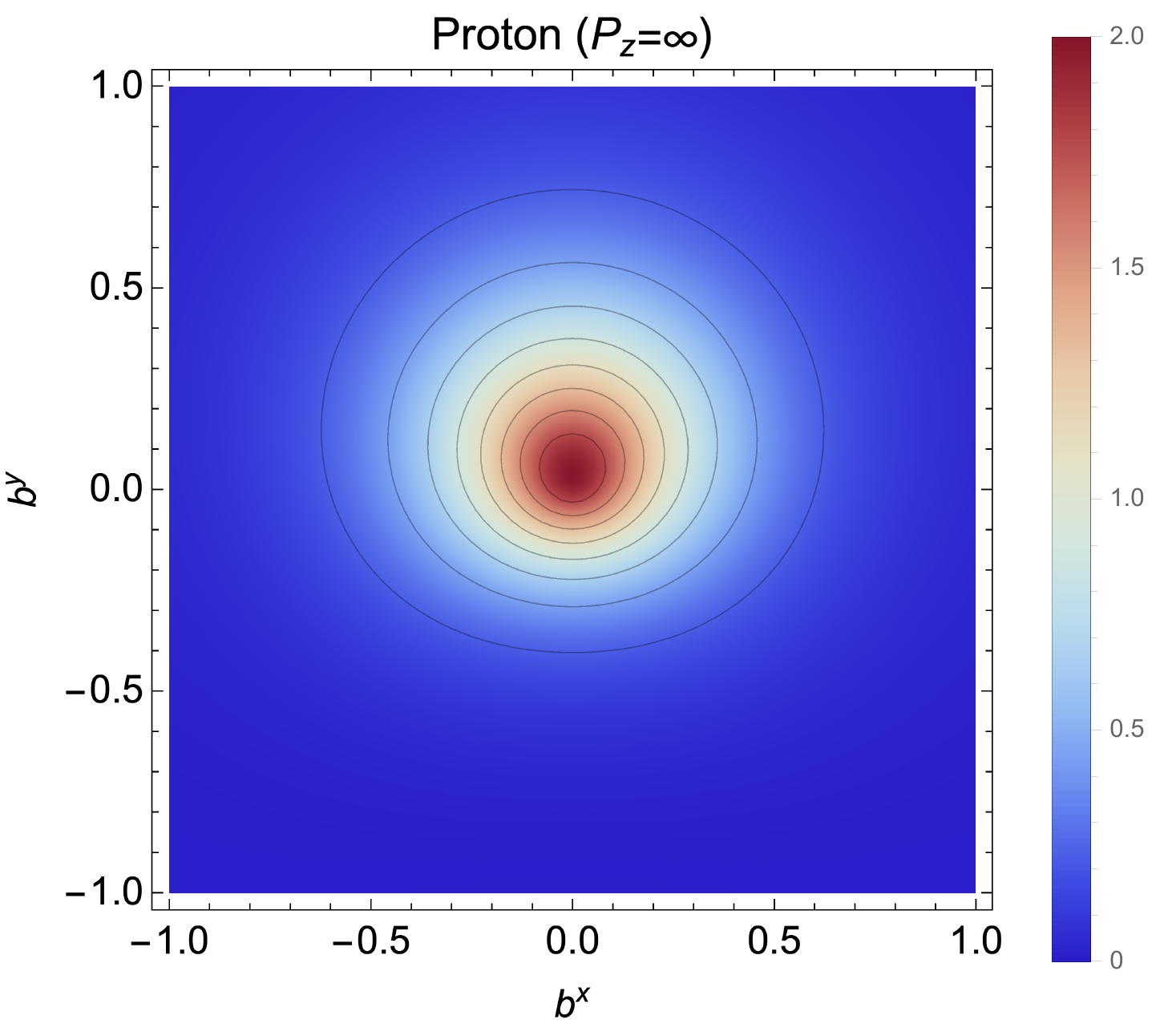}
\caption{The transverse charge distributions of the transversely
  polarized proton along the $b_x$-direction are illustrated in the 2D
  transverse plane with $P_{z}$ varied from zero to infinity.}  
\label{fig:5} 
\end{figure}
\begin{figure}[htp]
\includegraphics[scale=0.39]{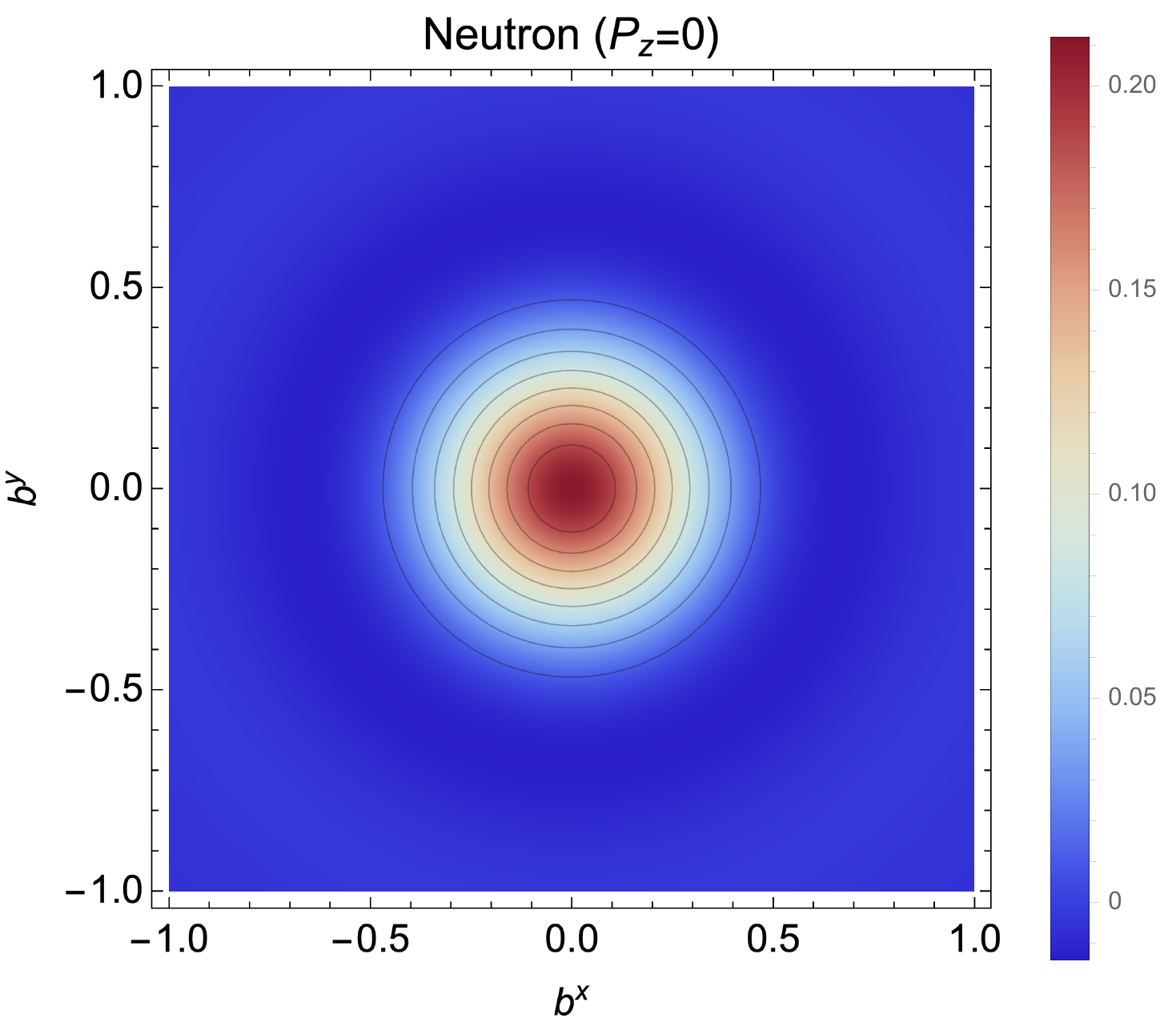}
\includegraphics[scale=0.39]{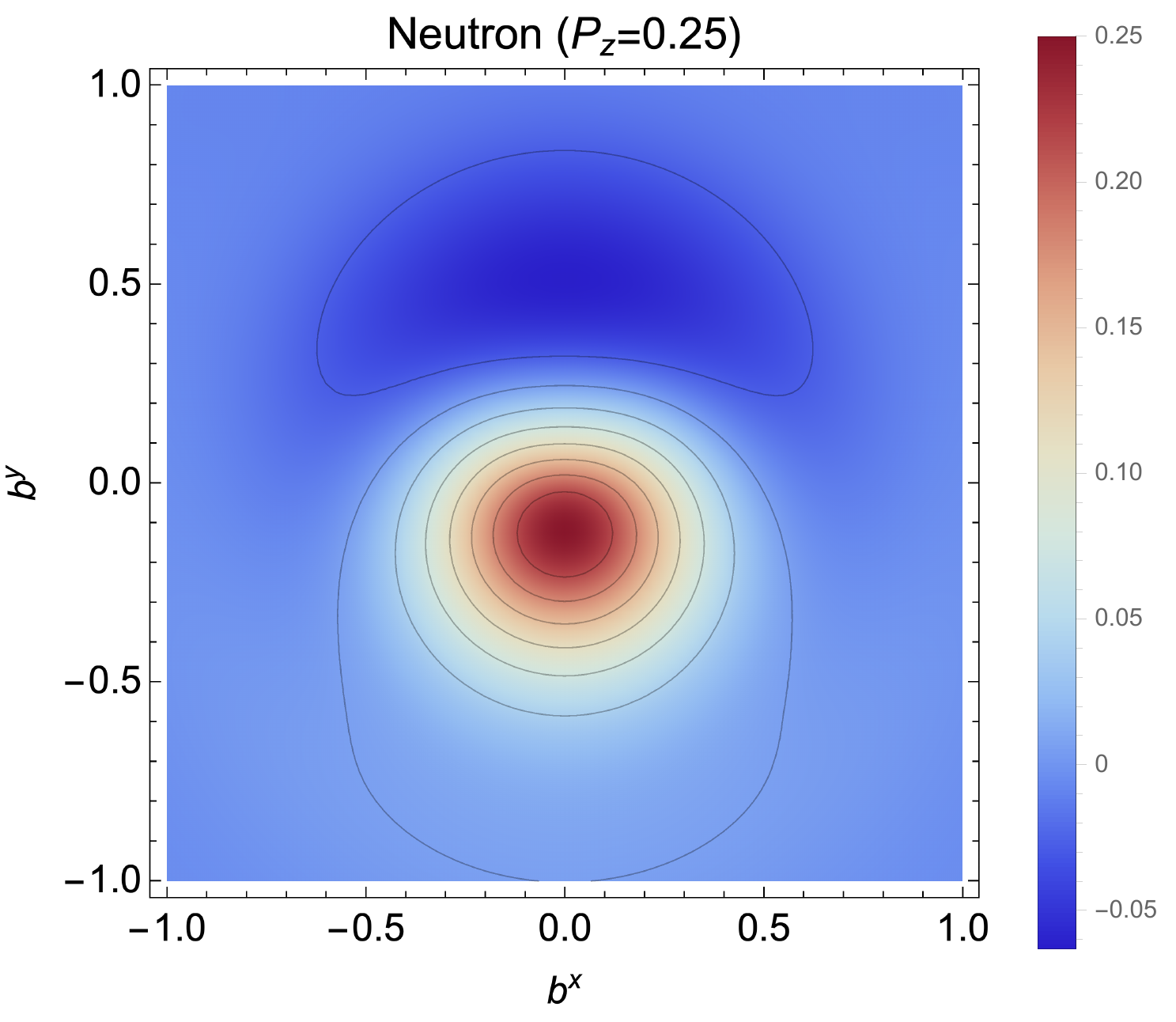}
\includegraphics[scale=0.39]{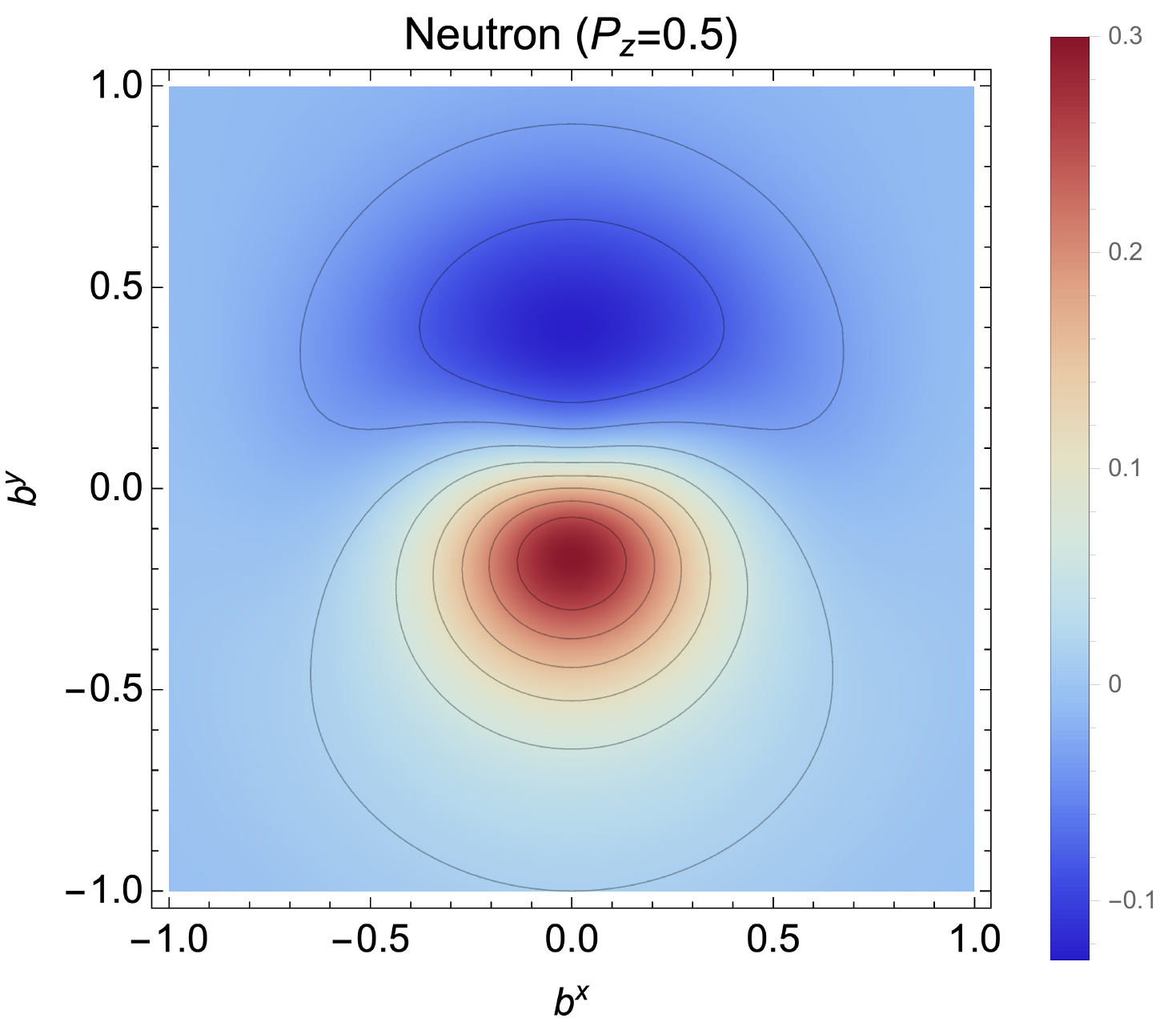}
\includegraphics[scale=0.39]{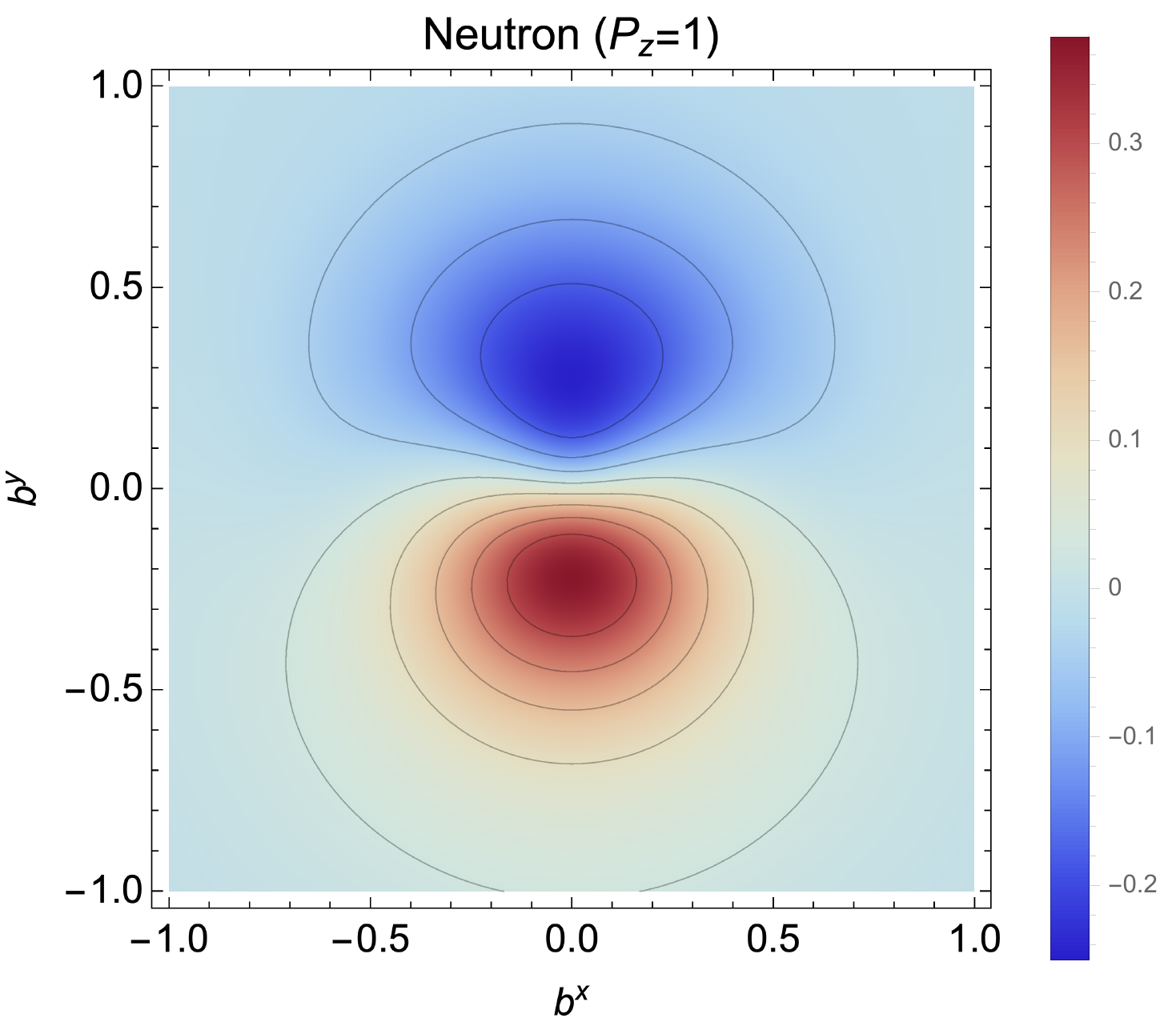}
\includegraphics[scale=0.39]{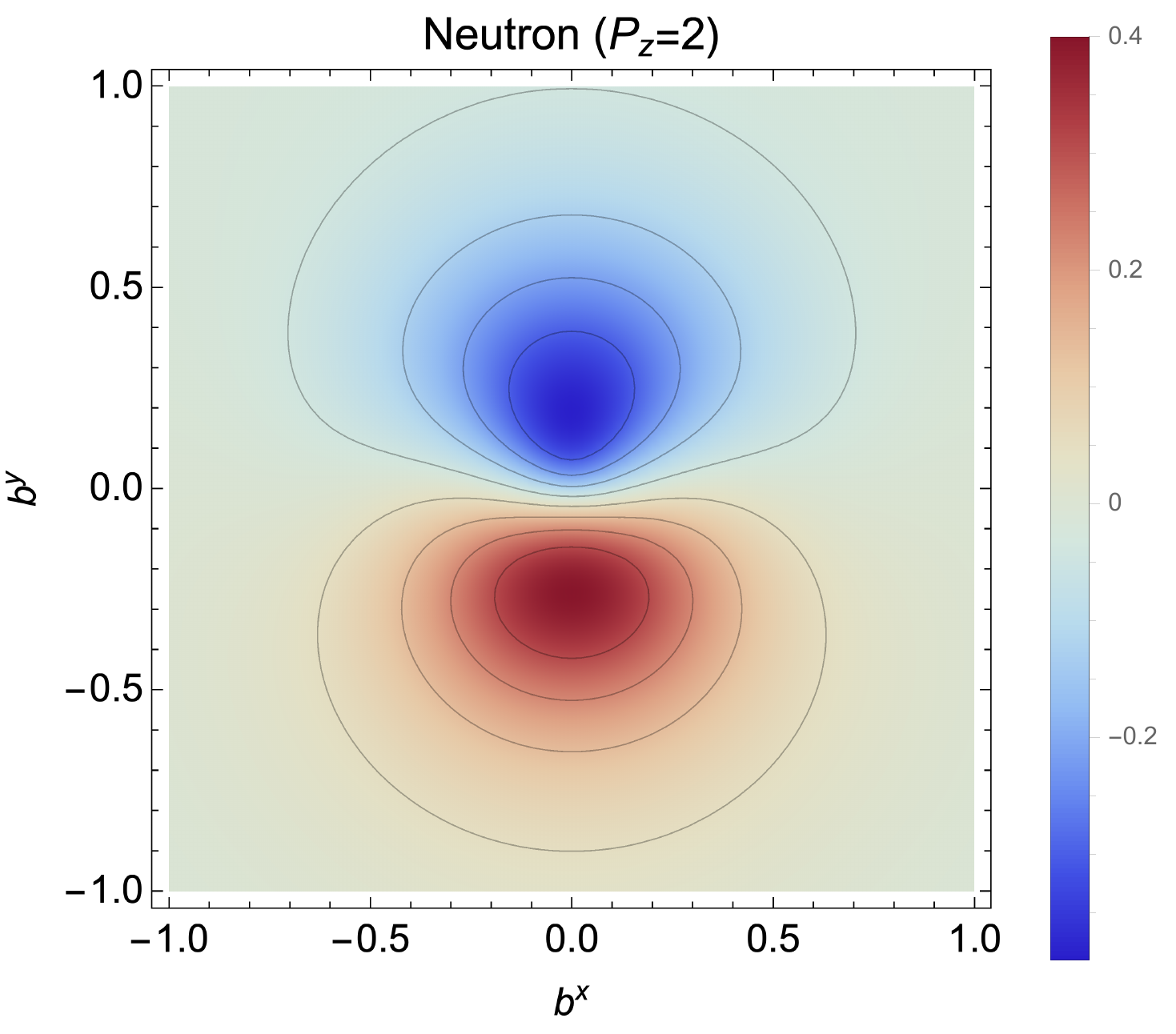}
\includegraphics[scale=0.39]{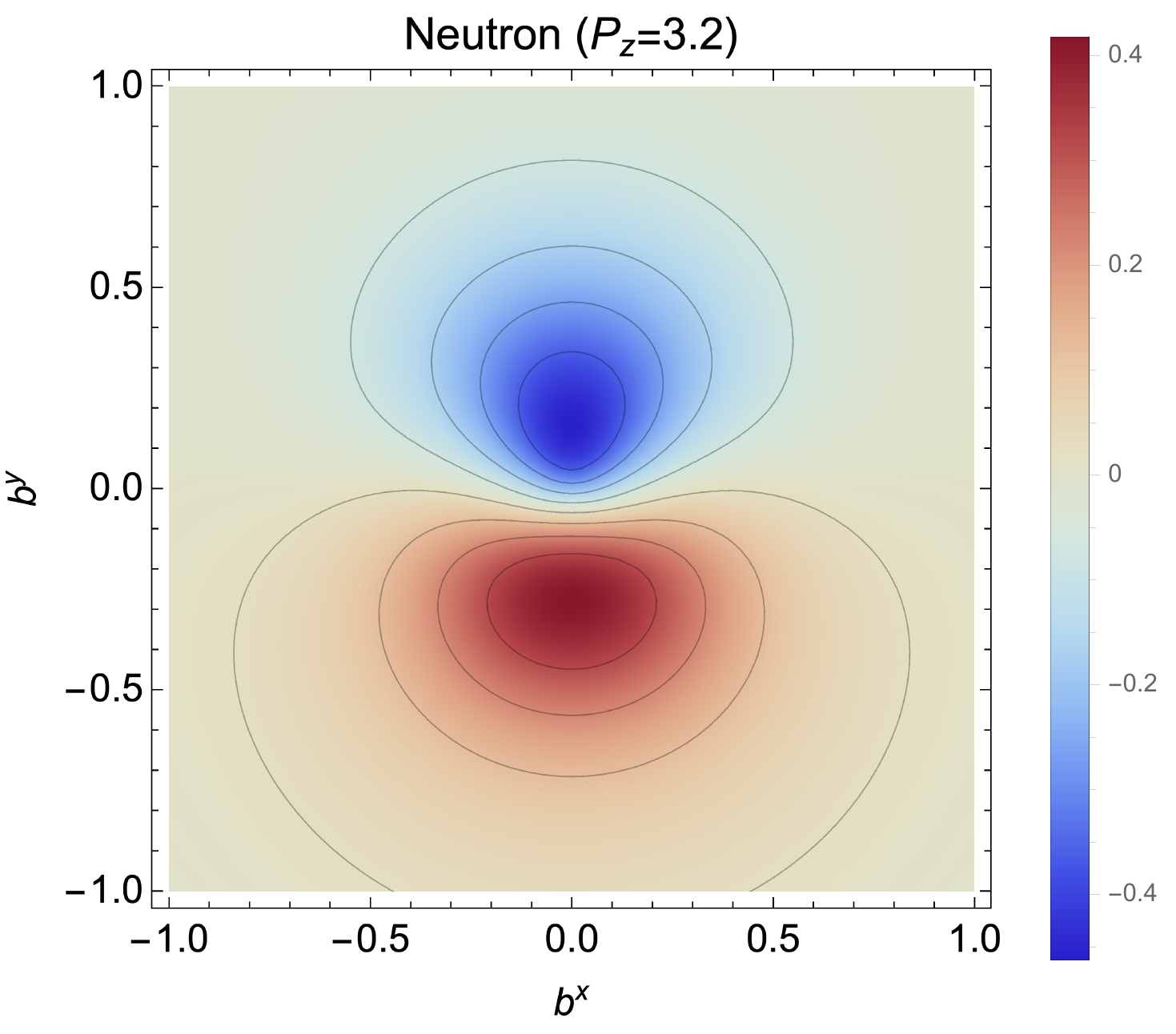}
\includegraphics[scale=0.39]{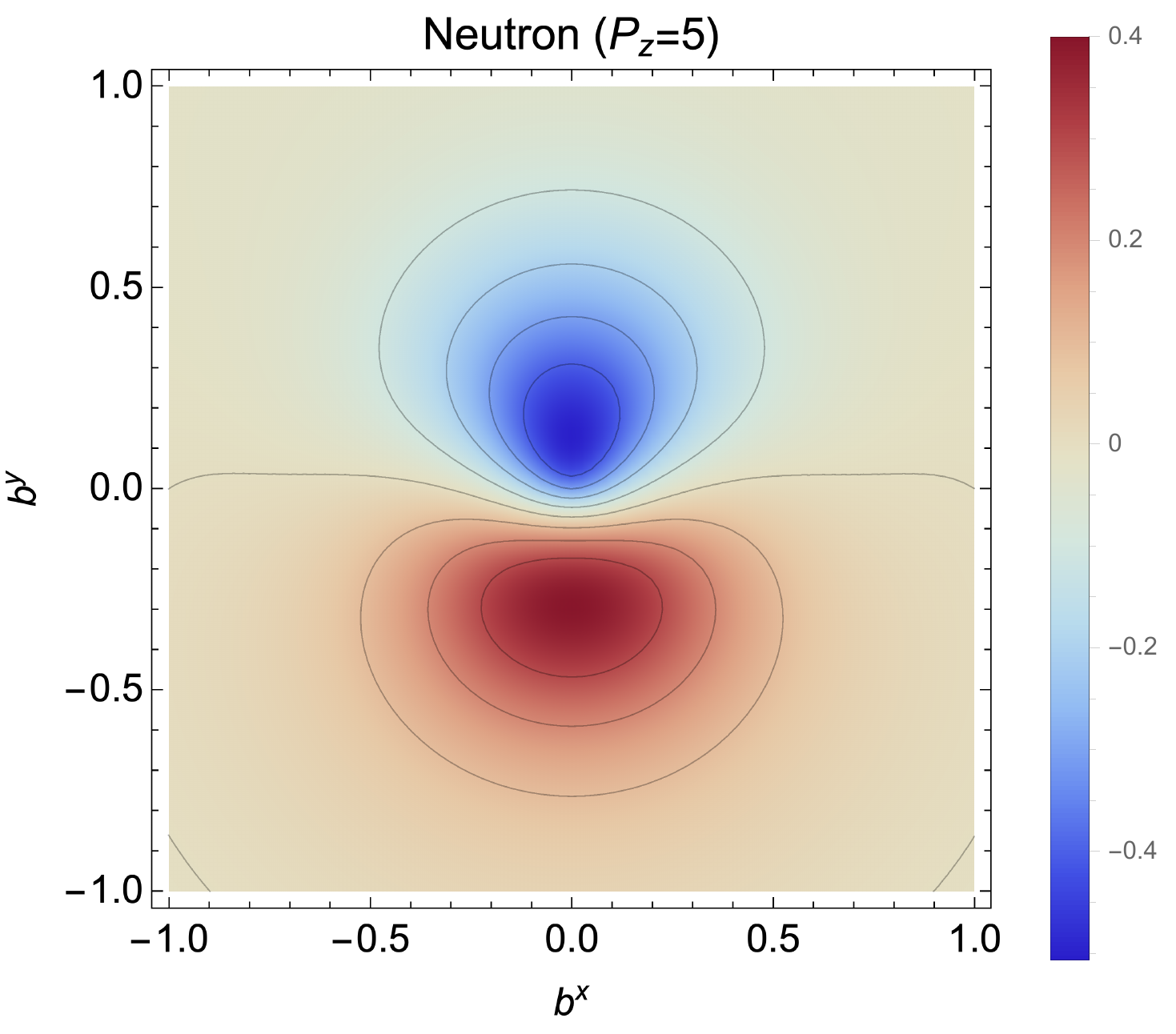}
\includegraphics[scale=0.39]{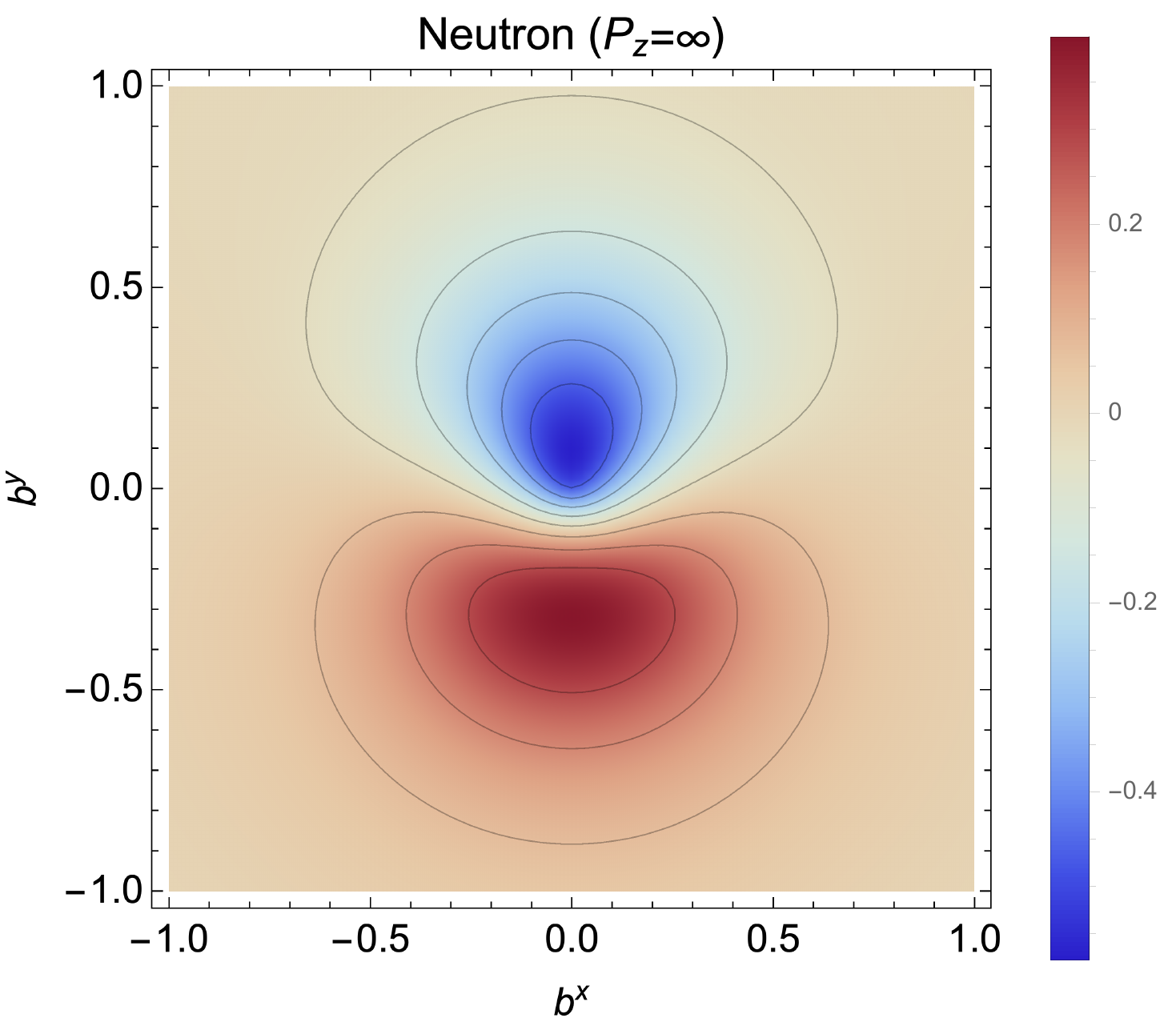}
\caption{The transverse charge distributions of the transversely
  polarized neutron along the $b_x$-direction are illustrated in the 2D
  transverse plane with $P_{z}$ varied from zero to infinity.}
\label{fig:6}
\end{figure}
When the nucleon is transversely polarized along the $b_x$-axis, its 
charge distribution starts to get deformed as $P_z$ increases. The
external magnetic field is exerted on the moving nucleon, 
the induced electric dipole moment, which is defined in
Eq.~\eqref{eq:elecdip}, brings about the induced electric field
$\bm{E}'=\gamma (\bm{v}\times \bm{B})$, where $\bm{v}$ is the velocity
of the moving nucleon in the EF. The solid curve in the left panel of
Fig.~\ref{fig:3} draws the magnitude of the nucleon induced electric
dipole moment as a function of $P_z$. The dashed line depicts the
value of $d_y$ in the IMF, which is identical to the proton anomalous
magnetic moment divided by $2M_N$, i.e., $\kappa_p/2M_N=0.188$ fm. As
expected, $d_y$ starts to increase as $P_z$ increases. Interestingly,
however, $d_z$ reaches its maximum value at around $P_z$ and then
diminishes gadually. As $P_z\to \infty$, it converges on the dashed
line. On the other hand, $d_y$ of the neutron becomes negative and its
magnitude grows monotonically as $P_z$ increases as illustrated in the
right panel of Fig.~\ref{fig:3}. It also converges on
$\kappa_n/2M_N=-0.201$ fm.  

The finite value of $d_y$ gives rise to the displacement of the charge 
distribution of the transversely polarized nucleon as shown in
Fig.~\ref{fig:1}~\footnote{In fact, the charge distributions of the 
transversely polarized nucleon in the IMF were studied in 
Ref.~\cite{Carlson:2007xd}. However, the directions of their
displacements are opposite to those in the present work.}. As $P_z$
increases, the left panel of Fig.~\ref{fig:4} shows how the charge
distribution $\rho_{\mathrm{ch}}^{T}(b, P_z)$ of the transversely polarized
proton is distorted. When the proton starts to move,
$\rho_{\mathrm{ch}}^{T}(b,P_z)$ of the proton begins to get displaced to the
positive $b_y$-direction. When $P_z$ reaches $3.2 \, \mathrm{GeV}$,
the proton charge distribution is maximally displaced and when it
reaches infinity the displacement of the distribution is slightly 
weakened. When it comes to the neutron case, the situation is more
interesting. As shown in the right panel of Fig.~\ref{fig:4}, the
charge distribution of the transversely  
polarized neutron starts to get deformed in a quite asymmetric way, as 
$P_z$ increases. The results indicate that the down quarks are
displaced to the positive $b_y$-direction whereas the up quark is
shifted to the negative side. In Figs.~\ref{fig:5} and~\ref{fig:6}, we
illustrate respectively the 2D transverse charge distributions of the
transversely polarized proton and neutron,varying $P_z$ from zero to
infinity. Figure~\ref{fig:5} shows explicitly how the proton
charge distribution $\rho_{\mathrm{ch}}^T(b,P_z)$ is distorted as
$P_z$ increases. We observe that $\rho_{\mathrm{ch}}^T(b,P_z)$ is
displaced along the positive $b_y$-direction, which is caused
by the induced electric dipole field. On the other hand,
$\rho_{\mathrm{ch}}^T(b,P_z)$ of the neutron is deformed in a more
prominent way. When the neutron is at rest, its transverse charge
distribution is shaped in a symmetric 
form. When it starts to move, the positive charges move to the
negative $b_y$-direction whereas the negative ones get displaced along
the positive $b_y$. This is due to the negative values of the electric
dipole moment. 

\section{Abel transformation between the 3D Breit and 2D Drell-Yan 
  frames \label{sec:3}}  
Recently, Ref.~\cite{Panteleeva:2021iip} showed that the 2D
light-front force distributions can be expressed in terms of the 3D BF
pressure and shear-force distributions, utilizing the Abel
transformation. This indicates that the Abel transformation can map
the 3D BF distributions onto the 2D ones in the IMF and they can be
reconstructed from the 2D ones by the inverse Abel transform.
So, the 3D distributions can still provide intuitive
information on the nucleon, though they can only be interpreted as
quasi-probabilistic ones. We now want to show how the Abel
transformations can be also used to map the 3D charge distributions
onto the 2D transverse ones in the IMF.     
\subsection{Formalism}
In the Wigner sense, the Breit frame can be viewed as the rest frame
($\bm{P}=\bm{0}$) of the nucleon localized around  $\bm{R}$. The
charge ($J^{0}$) and magnetic $(\nabla \times \bm{M})$
quasi-probabilistic distributions are given as the Fourier transforms 
of the Sachs electric and magnetic form factors, respectively: 
\begin{align}
\rho^{\mathrm{BF}}_{\mathrm{ch}}(r)= \int \frac{d^{3} \Delta}{(2\pi)^{3}}
  e^{-i \bm{x}\cdot \bm{\Delta}}  \frac{m}{P^{0}} G_{E}(t), \ \ \
  \rho^{\mathrm{BF}}_{\mathrm{M}}(r)= \int
  \frac{d^{3} \Delta}{(2\pi)^{3}} e^{-i \bm{x}\cdot \bm{\Delta}}
  \frac{m}{P^{0}}  G_{M}(t), 
\label{eq:BF}
\end{align}
which are the fully relativistic expressions~\cite{Lorce:2020onh}. 
If we take the nonrelativistic limit, i.e., ~${m}/{P^{0}}\sim 1$, we
arrive at the expressions
\begin{align}
\rho^{\mathrm{NR}}_{\mathrm{ch}}(r)= \int
  \frac{d^{3}\Delta}{(2\pi)^{3}} e^{-i \bm{x}\cdot \bm{\Delta}}
  G_{E}(t), \ \ \ \rho^{\mathrm{NR}}_{\mathrm{M}}(r)= \int
  \frac{d^{3}\Delta}{(2\pi)^{3}} e^{-i \bm{x}\cdot \bm{\Delta}}
  G_{M}(t), 
\label{eq:NRexp}
\end{align}
which can be often found in textbooks. However, the definitions of the
charge and magnetization distributions given in Eq.~\eqref{eq:NRexp}
have been criticized over decades~\cite{Yennie:1957, Burkardt:2000za,  
  Burkardt:2002hr, Miller:2007uy, Miller:2010nz, Jaffe:2020ebz,
  Lorce:2020onh}. As mentioned already, the main criticism comes from
the fact that the size of the nucleon is comparable to its Compton
wavelength, which implies that the nucleon is a relativistic
particle. So, we are not able to construct  
the wavepacket for the nucleon, which can be localized within the
space characterized by the Compton wavelength. Thus, the charge
distribution defined in Eq.~\eqref{eq:NRexp} cannot be interpreted as
the quantum-mechanical probabilistic one. 
To define the charge distribution correctly as a  
quantum-mechanical probabilistic one, one should introduce the
corresponding one in the transverse plane to the direction of the
moving nucleon with the infinite longitudinal
momentum~\cite{Burkardt:2000za, Burkardt:2002hr, Miller:2007uy,
  Miller:2010nz}. While this low-dimensional transverse charge
distribution is considered as a price one should pay to define it
as a quantum-mechanically meaningful one, it rather paves the way to
understand the internal structure of the nucleon in a more profound
way. Various transverse distributions, which correspond to the EM,
gravitational and tensor form factors, reveal unprecedented features
of the nucleon. In this sense, the Abel and Radon transformations also 
provide a new way of examining the structure of the nucleon. 

As mentioned in the previous section, the transverse charge and
magnetization densities are obtained respectively by the Fourier
transforms of the Dirac and Pauli form factors
\begin{align}
&\rho_{\mathrm{ch}}(b)= \int \frac{d^{2}\bm{\Delta}}{(2\pi)^{2}}
  e^{-i \bm{b}\cdot \bm{\Delta}}  F_{1}(t), \ \ \
\rho_{\mathrm{M}}(b)= \int  \frac{d^{2}\bm{\Delta}}{(2\pi)^{2}} 
e^{-i  \bm{b}\cdot \bm{\Delta}}  F_{2}(t).
\label{eq:distDYF}
\end{align}
Employing now the Abel transform~\cite{Abel, Natterer:2001} for a
spherically symmetric particle such as the nucleon,   
we can map the 3D charge and magnetization distributions onto the
transverse charge and magnetization distributions. In fact, the Abel  
transformation was already applied to the deeply virtual Compton
scattering~\cite{Polyakov:2007rv, Moiseeva:2008qd} and the
energy-momentum tensor distributions~\cite{Panteleeva:2021iip,
 Kim:2021jjf}. The Able transform and its inverse transform
are defined as  
\begin{align}
A[g](b) =\mathcal{G}(b) = \int^{\infty}_{b}
  \frac{dr}{r} \frac{g(r)}{\sqrt{r^{2}-b^{2}}}, \ \ \  
g(r) =  - \frac{2}{\pi} r^{2} \int^{\infty}_{r} d b
  \frac{d\mathcal{G}(b)}{db} 
  \frac{g(r)}{\sqrt{b^{2}-r^{2}}}.   
\label{eq:Able}
\end{align}
Thus, $A[g](b):= \mathcal{G}(b)$ is called the Abel image of the
function $g(r)$. In addition, a useful relation for the Mellin
moments of the Abel images can be obtained 
as~\cite{Panteleeva:2021iip}:  
\begin{align}
& \int^{\infty}_{0} b^{N}   A[g](b)\,db=\frac{\sqrt{\pi}}{2}
  \frac{\Gamma\left(\frac{N+1}{2}\right)}{ 
\Gamma\left(\frac{N+2}{2}\right)} 
  \int^{\infty}_{0}  \, r^{N-1} g(r)  dr.
\label{eq:Mel}
\end{align}
The relation~\eqref{eq:Mel} is valid as far as the 3D EM distributions
in the distance decrease faster than any order of
$r^{N-1}$. The 2D distribution in the IMF can be easily connected to
the 3D distribution in the BF 
\begin{align}
&\rho_{\mathrm{ch}}(b) +  \frac{1}{4M_N^{2}}\partial_\perp^{2}
  \rho_{\mathrm{M}}(b) =  \int^{\infty}_{b}
  \frac{2r\,dr}{\sqrt{r^{2}-b^2}} 
  \rho^{\mathrm{NR}}_{\mathrm{ch}}(r), \ \
  \rho_{\mathrm{ch}}(b) +  \rho_{M}(b) = \int^{\infty}_{b}
  \frac{2r\,dr}{\sqrt{r^{2}-b^2}} \rho^{\mathrm{NR}}_{\mathrm{M}}(r),  
\label{eq:abeltr2D}
\end{align}
which are just the Abel transforms of the 3D charge and magnetization
distributions . It is of great importance to observe that when the
Abel transforms of the 3D BF charge and magnetization distributions
yield naturally the combination of the transverse charge and
magnetization ones, which are the consequence of the Lorentz boost. In
fact, it is one of the essential features of the Abel transform that
it properly incorporates the effects of the 
Lorentz boost. 
 
Making use of Eq.~\eqref{eq:Able}, the Abel images of the 3D charge
and magnetic distributions in the BF are expressed by 
\begin{align}
&A[2r^{2}\rho^{\mathrm{NR}}_{\mathrm{M}}](b) = 
  \rho_{\mathrm{ch}}(b)+   \rho_{M}(b),\;\;\; 
  A[2r^{2}\rho^{\mathrm{NR}}_{\mathrm{ch}}(r)](b) = 
  \rho_{\mathrm{ch}}(b)+
  \frac{1}{4M_N^{2}}\partial_\perp^{2}\rho_{M}(b),
\label{eq:AbelI}
\end{align}
which are identical to the 2D transverse charge and magnetization
distributions. Note, however, that it is difficult to express the 3D
distributions together with the relativistic corrections $\sim
1/P^{0}$ in the BF in terms of the 2D transverse distributions in 
the IMF due to the non-locality of the kinematical factor $P^{0}$. It
is also interesting to examine the Mellin moment of the Abel
image. For $N=1$, one has an equality of the charge and magnetic
moments in both the 2D and 3D frames. For $N=3$, one arrives at
\begin{align}
&  Q \langle r^{2}
  \rangle_{\mathrm{ch}}  =  \frac{3}{2} \left(Q \langle b^{2}
  \rangle_{\mathrm{ch}}  +\frac{\kappa_N}{M_N^{2}}\right),
  \;\;\;  
\langle r^{2} \rangle_{M}  = \frac{3}{2}\frac{\langle
  b^{2} \rangle_{\mathrm{ch}} Q + \langle b^2 
  \rangle_{M}\kappa_N}{Q+\kappa_N}, 
\end{align}
where $Q$ represents the charge of the nucleon and $\kappa_N$ is the
corresponding anomalous magnetic moment. 

\subsection{Results and discussion}
\begin{figure}[htp]
\includegraphics[scale=0.285]{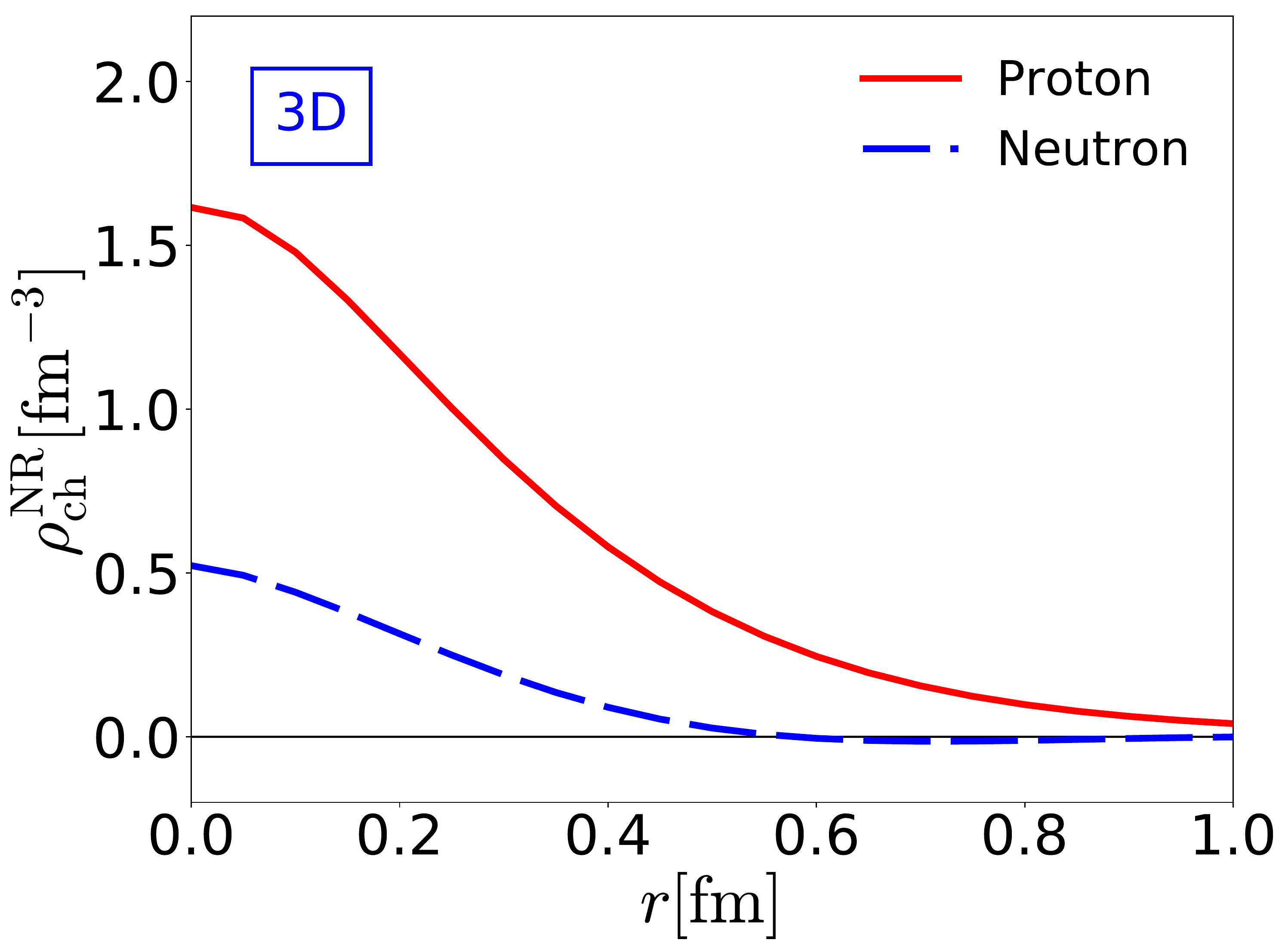}
\includegraphics[scale=0.285]{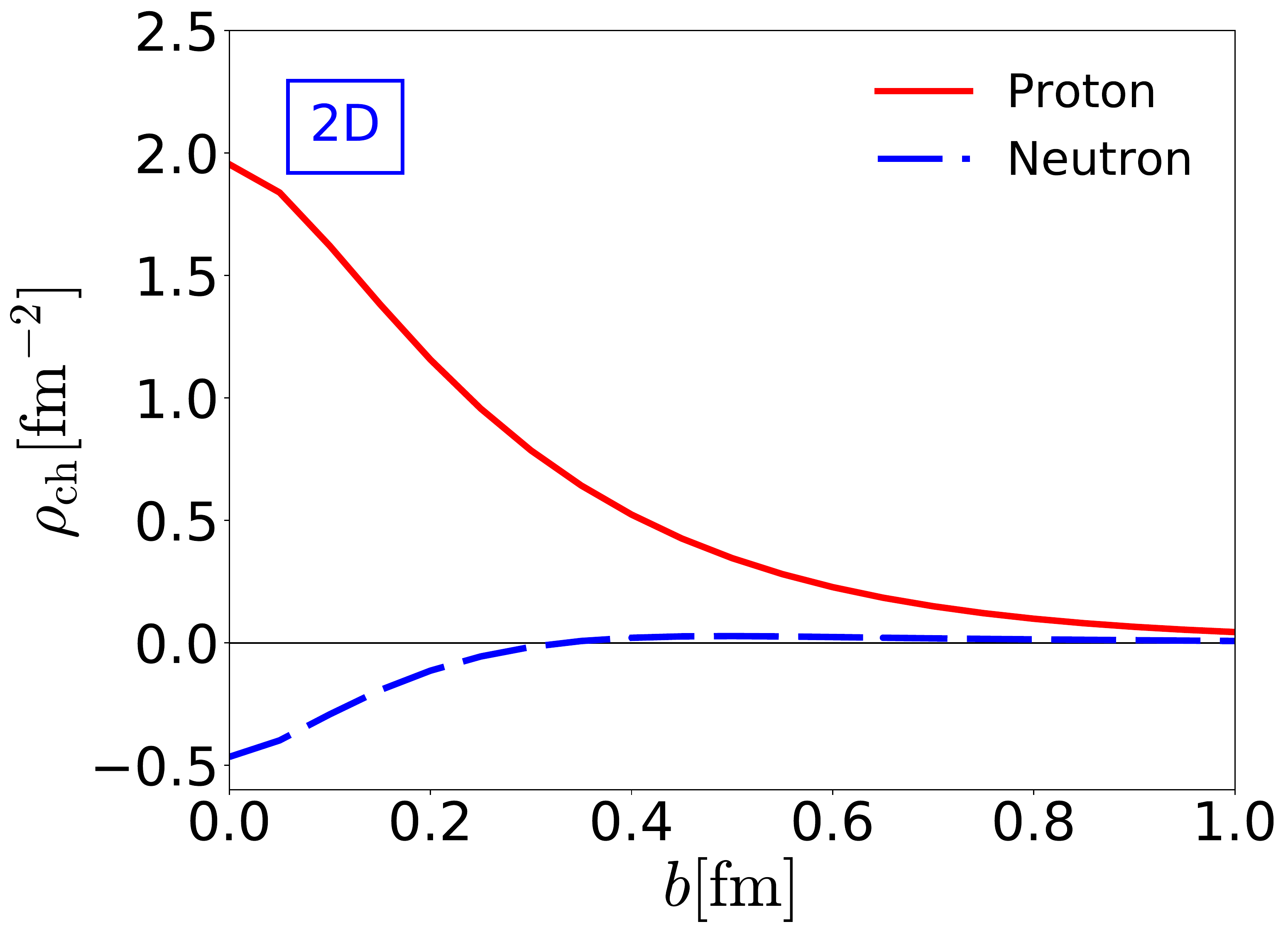}
\includegraphics[scale=0.285]{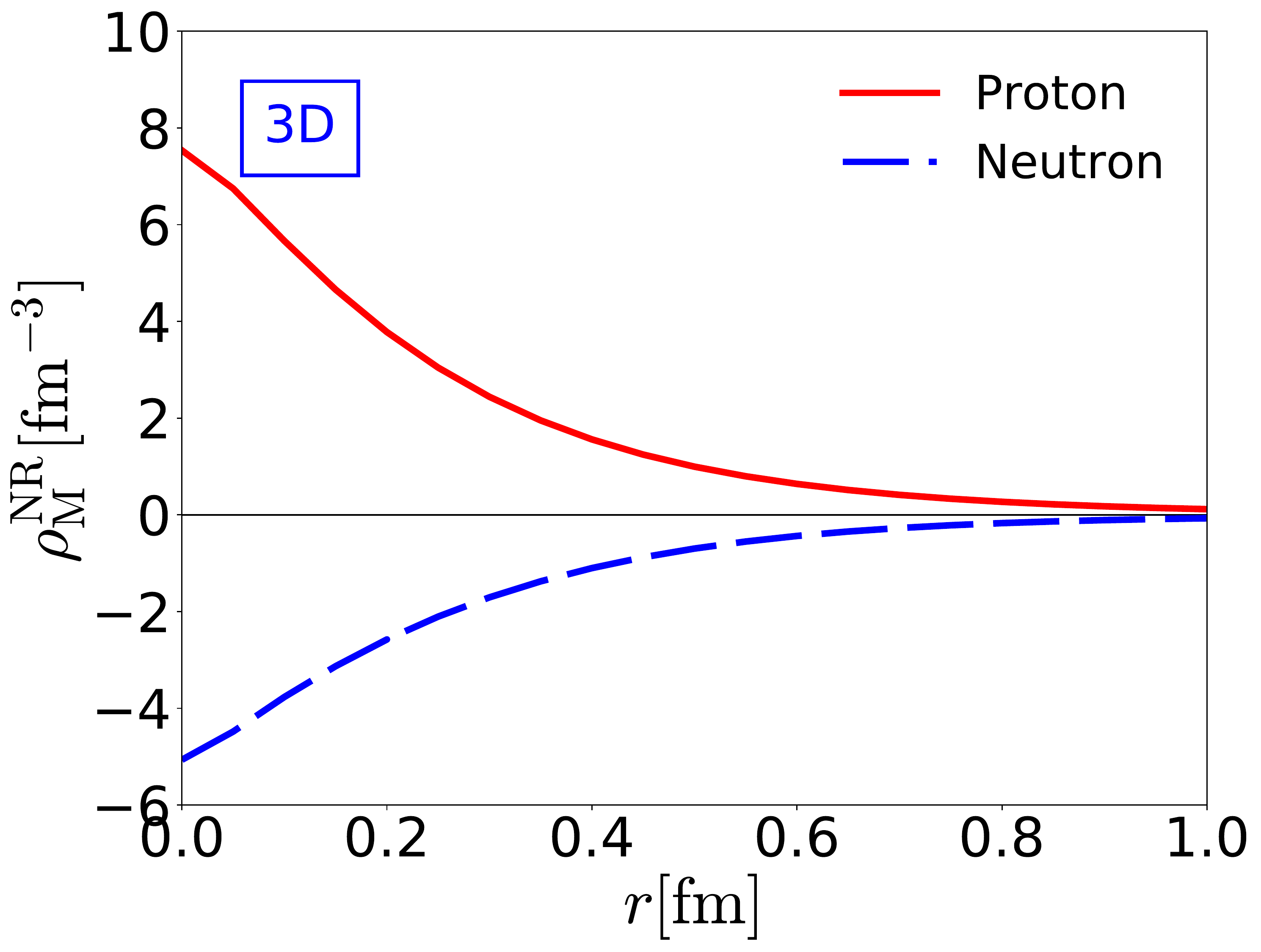}
\includegraphics[scale=0.285]{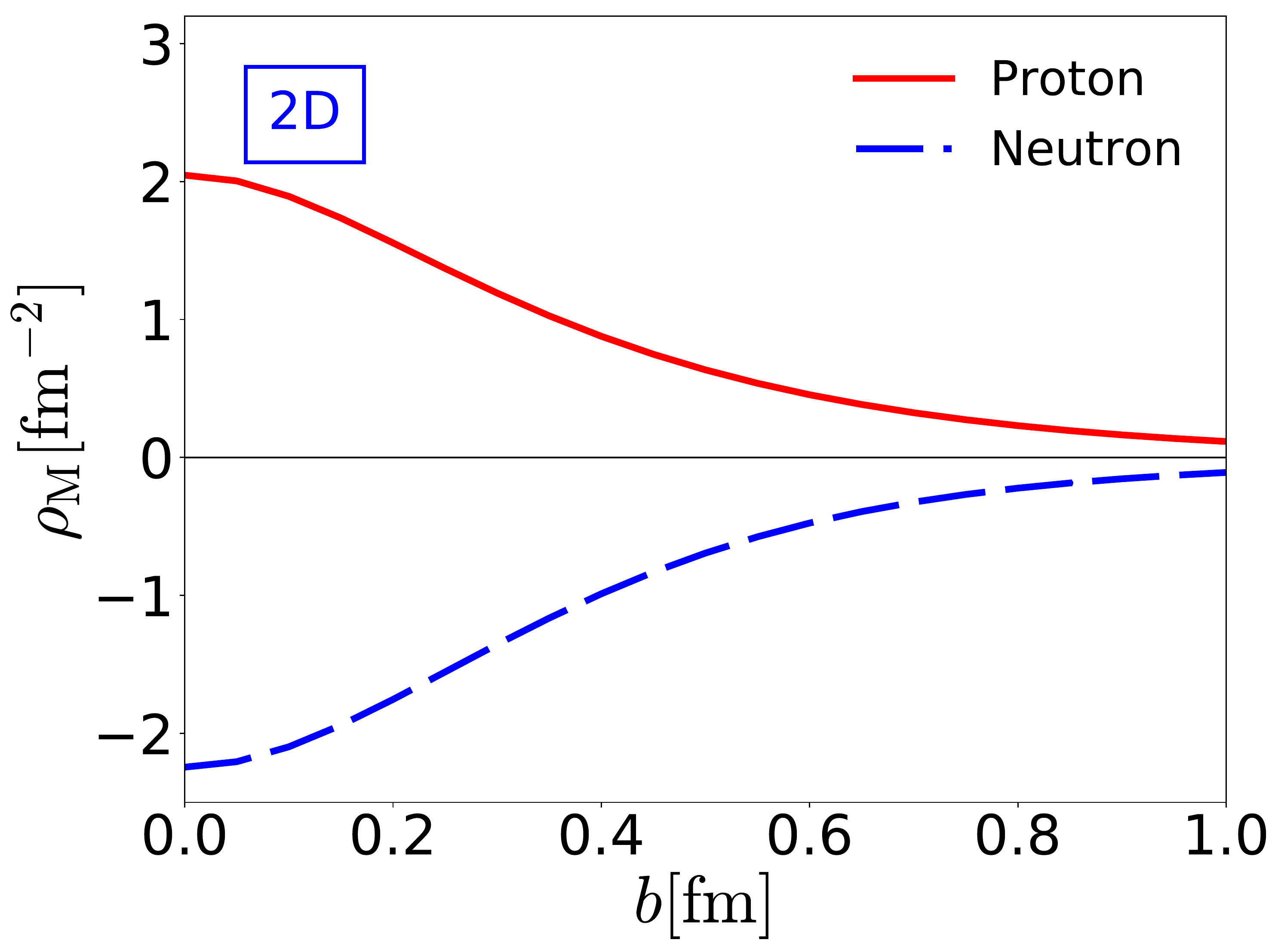}
\caption{The 3D (transverse) charge and magnetization distributions are
  presented in the BF (IMF) in the left (right) upper and lower
  panels, respectively. The solid and dashed curves denote the cases
  of the proton and neutron, respectively. The charge distributions are
  normalized to be its charge whereas the 3D (2D) magnetization
  distribution is normalized to be the corresponding magnetic dipole
  moments $\mu_{p}$ and $\mu_n$ in units of the nuclear magneton
  (anomalous magnetic moment $\kappa_{N}$).}  
\label{fig:7} 
\end{figure}
The upper panel of Figure~\ref{fig:7} draws the 3D charge distributions
in the BF (left panel) and the transverse charge distributions in the
IMF (right panel). As anticipated, both the transverse charge
distributions of the proton and neutron are properly obtained from the
3D ones by the Abel transformation. The results for the 2D transverse
charge distributions are identical to those obtained in
Fig.~\ref{fig:1} when $P_z\to \infty$. 
In particular, it is remarkable to see that the transverse charge
distributions of the neutron in the IMF are correctly
reproduced. As already found in Eq.~\eqref{eq:abeltr2D}, As $P_z\to
\infty$, the effect of the magnetization enters into the transverse
charge distribution, which makes core part of the neutron turn 
negative. Thus, the Abel transformation maps correctly the 3D charge
distributions of the nucleon at rest onto the 2D transverse ones in
the IMF.

As shown in the lower panel of Fig.~\ref{fig:7}, both the
transverse magnetization distributions of the proton and neutron are
also derived correctly, which have also shown in Fig.~\ref{fig:2}. The
present results draw two important physical implications: Firstly, the
Abel transformation can also be applied to the charge and magnetization 
distributions of the nucleon. Secondly, while the 3D charge
distributions have only the quasi-probabilistic meaning in the Wigner
sense, they still give us valuable insight into understanding the
internal structure of the nucleon with the help of the Abel
transformations.   

\section{Summary and conclusions~\label{sec:4}}
In the present work, we aimed at investigating how the transverse
charge and magnetization distributions of both the unpolarized and
transversely polarized nucleon get displaced as its longitudinal
momentum increases from zero to infinity. Interestingly, the
transverse magnetization distribution of the nucleon is maximized not
at the infinite momentum but at around 3.2 GeV. This originates from
the kinematical reason. When the nucleon is transversely polarized
along the positive $x$-direction of the impact parameter, the induced
electric dipole moment is produced at the finite value of the
longitudinal momentum. This causes the displacement of the transverse
charge distribution of the proton along the positive $y$-direction. On
the other hand, that of the neutron shows more interesting
behavior. The positive charges, which represent the up valence quark
inside a neutron, are displaced to the negative direction whereas the
negative charges or the down valence quarks are moved toward the
positive direction. This is due to the negative values of the electric
dipole moment of the neutron. 

In the second part of the present work, we examined how the Abel
transformations map the 3D charge distributions in the Breit frame
on to the 2D transverse ones in the Drell-Yan frame or the infinite
momentum frame. As expected, the Abel transformations successfully
project the 3D BF charge distributions onto the 2D transverse plane
in the Drell-Yan frame. Both the 3D charge and magnetization
distributions in the Breit frame are expressed in terms of the
transverse charge and magnetization ones in the Drell-Yan frame. Keep
in mind that the Lorentz boost makes the charge and magnetization 
distributions mixed, the Abel transformations respect the Lorentz
boost, so that we have properly reproduced the transverse charge
distributions in the Drell-Yan frame. We found the physical
implications of the Abel transformation in the present work: The Abel
transformation indeed connects the 3D charge and magnetization
distributions to the 2D transverse ones, which implies that the 3D
distributions can still provide insight into the internal structure of
the nucleon, even though they have only the quasi-probabilistic
meaning.  

When we consider the interpolation between the 3D distributions and 2D
transverse ones for particles with spin higher than 1/2, one has to
employ the Radon transformation instead of the Abel one, as already
proposed in~\cite{Panteleeva:2021iip}. The corresponding investigation
is under way.

%-------------------------------------------------
\begin{acknowledgments}
%-------------------------------------------------
The authors are very grateful to M. V. Polyakov for invaluable
discussions and suggestions. The present work was supported by Basic
Science Research Program through the National Research Foundation of
Korea funded by the Ministry of Education, Science and Technology 
(Grant-No. 2018R1A5A1025563). J.-Y.K is supported 
by the Deutscher Akademischer Austauschdienst(DAAD) doctoral
scholarship and in part by BMBF (Grant No. 05P18PCFP1). 
\end{acknowledgments}

\end{document}